\documentclass[structabstract]{aa}  
\usepackage{natbib}
\usepackage{array}
\bibpunct{(}{)}{;}{a}{}{,}
\usepackage[fleqn]{amsmath}
\usepackage{amssymb}
\usepackage[dvips]{graphicx}
\usepackage{slashbox}
\usepackage{xfrac}
\usepackage{breakurl}
\usepackage{txfonts}
\usepackage{breqn}

\usepackage{pstricks}
\usepackage{multirow}
\usepackage{caption}
\usepackage{subcaption}
\usepackage{stfloats}
\usepackage{float}
\usepackage{morefloats}
\usepackage{color}
\usepackage{breqn}
\usepackage[shortcuts]{extdash}
\newcommand{\mic}{$\mu$m}
\newcommand{\Conv}{\mathop{\scalebox{1.5}{\raisebox{-0.2ex}{$\ast$}}}}%

\begin{document}
   \title{Halo dust detection around NGC 891\thanks{{\it Herschel} is an ESA space observatory 
   with science instruments provided by European-led Principal Investigator consortia and with important 
   participation from NASA.}}

   \author{M. Bocchio\inst{1, 2}, 
          S. Bianchi\inst{1},
          L. K. Hunt\inst{1},
          R. Schneider\inst{2}}
          
   \institute{INAF - Osservatorio Astrofisico di Arcetri, Largo Enrico Fermi 5, 50125 Firenze, Italy\\
   	INAF - Osservatorio Astronomico di Roma, Via di Frascati 33, I-00040 Monteporzio, Italy\\
                          }
 
  \abstract
  % context heading (optional)
   {Observations of edge-on galaxies allow us to investigate the vertical extent and properties 
   of dust, gas and stellar distributions.
NGC 891 has been studied for decades and represents one of the best studied cases of an edge-on galaxy.}
  % aims heading (mandatory)
   {We use deep PACS data together with IRAC, MIPS and SPIRE data to study the vertical extent of dust emission
   around NGC 891. We also test the presence of a more extended, thick dust component.}
 % methods heading (mandatory)
 {By performing a convolution of an intrinsic vertical profile emission with each instrument PSF and comparing it with
 observations we derived the
scaleheight of a thin and thick dust disc component.}
  % results heading (mandatory)
{For all wavelengths considered the emission is best fit with the sum of a thin and a thick dust component.
The scaleheight of both dust components shows a gradient passing from 70 $\mu$m to 250 $\mu$m.
This could  be due to a drop in dust heating (and thus dust temperature) with the distance from the plane, or to a 
sizable contribution ($\sim 15 - 80\%$) of an unresolved thin disc of hotter dust to the observed surface brightness at shorter wavelengths. 
% possibly
%indicating a drop in dust temperature within the vertical extent.
The scaleheight of the thick dust component, using observations from 70 $\mu$m to 250 $\mu$m 
has been estimated to be $(1.44\pm 0.12)$\,kpc, consistent with previous estimates 
(extinction and scattering in optical bands and MIR emission).
%The thick dust component presents a scaleheight of $(1.44\pm 0.12)$\,kpc,
 The amount of dust mass at distances larger than $\sim 2$\,kpc 
from the midplane represents $2 - 3.3$\% of the total galactic dust mass and 
the relative abundance of small grains with respect to large grains is almost halved comparing to that
in the midplane.}
    % conclusions heading (optional), leave it empty if necessary 
{The paucity of small grains high above the midplane might indicate 
that dust is hit by interstellar shocks or galactic fountains and entrained together with gas.
The halo dust component is likely to be embedded in an atomic / molecular gas and heated by a 
thick stellar disc.}

%two dust component
%
%dust in the thin disc show a gradient in scaleheights. This can be either because of a gradient in
%temperature or because of contamination from ...
%
%dust in the thick halo has a scaleheight of ... and the mass in the halo is...
%
%It can be heated either from a thick stellar component or, if we admit the presence of
%contamination, can be heated by the galactic stellar radiation.
%
%the relative abundance of small grains drops off...


   \maketitle
%
%________________________________________________________________

\section{Introduction}

Nearby edge-on spiral galaxies offer a special opportunity to learn about the vertical structure and properties 
of dust, gas and stellar distributions that extend out from the galaxy midplane.
Several galaxies of this class have been observed in optical bands and show many common properties.
%Dust often shows as an evident dust lane, making these objects unique to study dust emission and
%extinction at the same time.
The high inclination of these galaxies makes them particularly well-suited to conduct 
multi-wavelength studies of the vertical structure of their disc.
Fitting optical images with radiative transfer models allows information on the dust and stellar distributions
to be obtained.
Stars and dust are found to follow an exponential distribution both radially and vertically: the dust
radial scalelength is $\sim 1.4$ times larger than that of the stellar distribution while its 
vertical scaleheight is about half of that of stars 
(\citealt{1987ApJ...317..637K,1997A&A...325..135X,1998A&A...331..894X,1999A&A...344..868X,
2004A&A...425..109A,2007A&A...471..765B,2010A&A...518L..39B,2012MNRAS.427.2797D,2014MNRAS.441..869D}).

Extending studies further out to galactic haloes, there is increasing evidence for 
the presence of dusty clouds up to a few kpc from the galactic disc that seem to be
related to the disc-halo cycle (e.g. \citealt{2012EAS....56..291H}).
Direct optical imaging of edge-on galaxies often shows extraplanar filaments of dense clouds 
backlit from stars (\citealt{1997AJ....114.2463H, 1999AJ....117.2077H, 2000AJ....119..644H,2004AJ....128..674R,
2004AJ....128..662T,2005ASPC..331..287H}).
An estimate of the mass of dust contained in these clouds indicate that a relevant fraction
of dust can be lifted up to these vertical distances without being destroyed \citep{2005ASPC..331..287H}. 

Moving even further from the galaxy main body, \cite{1994AJ....108.1619Z} observed dust extinction up 
to 200 kpc out in the galactic haloes of two nearby galaxies, NGC 2835 and NGC 3521.
This result was supported by \cite{2010MNRAS.405.1025M}, who 
measured the extinction of background quasars correlating to local galaxies and found 
evidence for dust in the intergalactic medium (IGM) up to 1 Mpc from galactic centres.

{\it GALEX} and {\it Swift} observations in UV bands by \cite{2014ApJ...789..131H} 
reveal diffuse UV light around late-type 
galaxies out to 5-20 kpc from the galactic disc.
The emission is rather blue and not consistent with light from extraplanar stars; the favoured
hypothesis is that this emission escapes from the galactic disc and 
scatters off dust in the halo.
This finding is consistent with the extinction measurements from \cite{2010MNRAS.405.1025M}, therefore
corroborating the evidence for dust in galactic haloes.

With the advent of {\it Spitzer} and more recently of {\it Herschel} (\citealt{2010A&A...518L...1P}) 
we have the necessary sensitivity to directly 
observe extraplanar dust thermal emission.
Aromatic bands have been observed as far as 6 kpc above the disc of NGC 5907 
(\citealt{2006A&A...445..123I}) and in galactic winds around several nearby sources (\citealt{2013ApJ...774..126M}). 
Continuum dust emission up to a similar height was found in NGC 891 by \cite{2007ApJ...668..918B}.

In this paper we focus on the vertical dust distribution in the well-known edge-on galaxy NGC 891.
This galaxy is at a distance of 9.6 Mpc (e.g. \citealt{2004ApJS..151..193S})
from us and it has an inclination of $\simeq  89.7^{\circ}$ (\citealt{1999A&A...344..868X}).
The dust distribution follows an exponential profile with a scalelength of $h_{\rm d} \sim 8$\,kpc
and a scaleheight of $z_{\rm d} \sim 0.3$\,kpc, thinner than the stellar distribution but more radially extended
 (\citealt{1999A&A...344..868X}).
Radiative transfer models by \cite{2008A&A...490..461B} and 
\cite{2000A&A...362..138P,2011A&A...527A.109P}, starting from
models by \cite{1999A&A...344..868X}, are able to reproduce the dust SED; thus the radiation field 
across the galaxy is reliably modelled. 
The galactic stellar distribution comprises the two classical components,
an old stellar bulge and an old stellar disc, with the addition of a clumpy component confined in a very thin 
region at the centre of the main disc.
\cite{2011A&A...531L..11B}, using the results from the 3D radiative transfer model from \cite{2008A&A...490..461B}
are able to reproduce most of the characteristics of far-infrared (FIR) and submm images obtained with 
the SPIRE instrument onboard {\it Herschel}.

\begin{figure*}[ht]
\begin{center}
% this image is produced by ..../profile/plot_images.pro
\includegraphics[trim=0mm 0mm 0mm 0mm,scale = 2.]{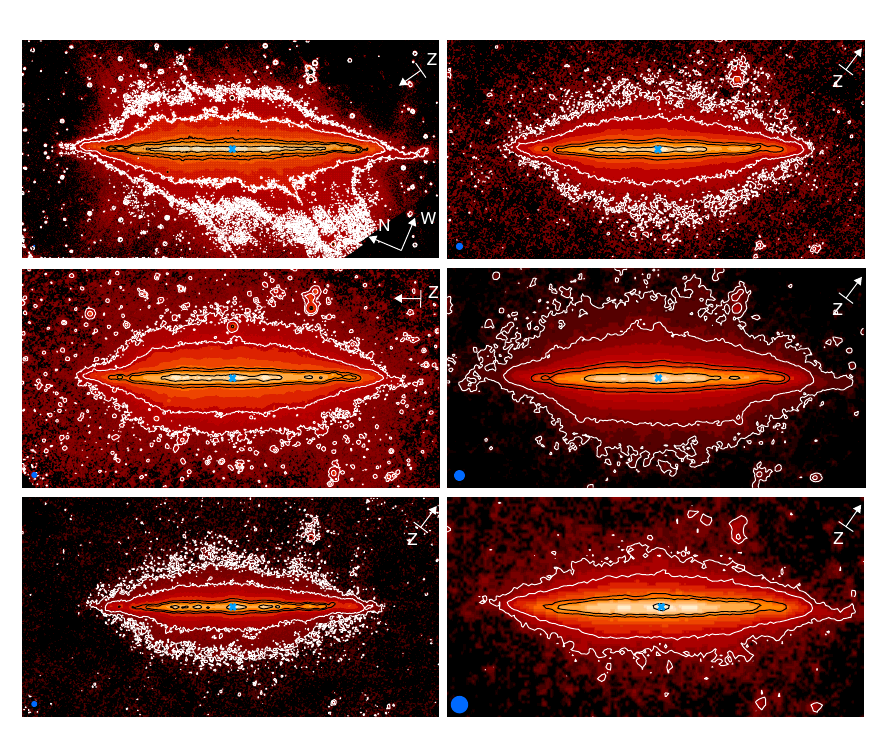}
\caption{From top to bottom: IRAC 8 $\mu$m (after stellar subtraction), 
MIPS 24 $\mu$m, PACS 70 $\mu$m (left column) and PACS 100 and 160 $\mu$m
and SPIRE 250 $\mu$m (right column) images of NGC 891.
Each panel shows an area of $\sim 14'$ x $7'$ centred on 
the galaxy's radio continuum coordinates $\alpha$ =
2$^{\rm h}$22$^{\rm m}$33$\fs$0, $\delta$ = 42$^{\circ}$20$'$57$\farcs$2 (J2000.0, blue cross; Oosterloo et al. 2007).
The main beam size is indicated by the filled blue circle (FWHM: $2''$, $6''$, $6''$, 
$7\farcs2$, $11''$, $17\farcs8$) and images have pixels sizes of $0\farcs6$,$1\farcs5$, $1\farcs5$, $2''$, $3''$, $6''$.
The white arrow at the top-right corner shows the Z-axis 
(i.e. the axis parallel to the short side of the instrument field of view) 
of the spacecraft during observations.
The sky RMS noise is $\sigma = 0.03, 0.03, 1.06, 1.23, 0.78$ and 1.31 MJy sr$^{-1}$. 
Contours are shown at 3, 10, 100, 200 
and 1000-$\sigma$ over all images. North and West directions are indicated.}
\label{fig:n891}
\end{center}
\end{figure*}

Optical broad-band observations obtained by \cite{1997AJ....114.2463H} with the WIYN 3.5-m telescope 
revealed the presence of a complex dusty structure above the galactic plane up to a distance $z \lesssim 2$\,kpc.
Dust emission was later directly detected by \cite{2007ApJ...668..918B}
with the Infrared Spectrograph aboard {\it Spitzer} at 16 and 22 \mic.
In particular, the surface brightness at 22 \mic\,drops following an exponential decay with a scaleheight of 
$1.3\pm0.3$\,kpc, consistently with the optical measurements by \cite{1997AJ....114.2463H}.
Furthermore, \cite{2007A&A...471L...1K} using MIPS 24 $\mu$m observations detected dust emission 
up to $\sim 2$\,kpc
from the galactic plane and \cite{2009MNRAS.395...97W}, using multi-wavelength observations, found 
extraplanar dust emission and aromatic features up to a vertical distance of $z\gtrsim 2.5$\,kpc.
However, \cite{2007A&A...471L...1K} and \cite{2009MNRAS.395...97W} did not consider in 
detail the effects of the instruments PSF and therefore their measurements are to be 
considered qualitative.

%- Kamphius et al. asymmetry in Halpha due to dust attenuation
More recently, \cite{2012IAUS..284..135S} reported on the discovery of a 
UV halo due to the diffuse dust present above the galactic plane of NGC 891 and
\cite{2014ApJ...785L..18S}, using a 3D radiative transfer model, found
that the UV halo is well reproduced by a model with two exponential dust discs, one
with a scaleheight of 0.2 - 0.25 kpc and a wider component with a scaleheight of 1.2 - 2.0 kpc.
%- Hughes et al. 2014? Dust temperature correlates with MIPS 24, estimate of dust mass. Maybe not necessary here

In this work, we use {\it Spitzer} and {\it Herschel} data (from 8 to 250 $\mu$m)
to study the extent of dust emission above the galactic plane,
taking a particular care in the analysis of the effects of the instruments {\em Point Spread Function} (PSF). 
We then compare the derived dust distribution to the extinction and emission
measurements from the literature.
This paper is organised as follows:
Section~\ref{sect:obs} presents the observations and the data reduction used to retrieve the data;
Section~\ref{sect:dust_em} shows the dust emission profile as measured by {\it Spitzer} and {\it Herschel}
instruments. 
In Section~\ref{sect:vert_var} we highlight the variability of the vertical profile  
along the major axis of the galaxy; Section~\ref{sect:dust_temp} shows measurements of the dust 
Spectral Energy Distribution, temperature and optical depth at different vertical distances. 
In Section~\ref{sect:discussion} we discuss our results and in Section~\ref{sect:summary} 
we draw our conclusions.

\section{Observations and data reduction}
\label{sect:obs}

{\it Herschel} PACS (\citealt{2010A&A...518L...2P}) photometric observations were taken at 
70, 100 and 160 \mic\,\,as part 
of the {\it Herschel} EDGE-on galaxy Survey (HEDGES;
PI: E. Murphy). 70 and 100 \mic\,\,data consist of two cross-scans at a $20''$ s$^{-1}$ scan speed 
(Obs IDs: 1342261790, 1342261792 for 70 \mic\,\,and
Obs IDs: 1342261791, 1342261793 for 100 \mic), while 160 \mic\,\,observations consist of four 
cross-scans at $20''$ s$^{-1}$ scan speed 
(Obs IDs: 1342261790, 1342261791, 1342261792, 1342261793).
The {\it Herschel} Interactive Processing Environment (HIPE, v.12.1.0;
\citealt{2010ASPC..434..139O}) was first used to bring the raw Level-0 data to Level-1 using the PACS
calibration tree PACS\_CAL\_65\_0. Maps were then produced 
using Scanamorphos (v.24.0, Roussel 2013).
Images were produced with pixel sizes of $1\farcs5$, $2''$ and $3''$, respectively (Fig.~\ref{fig:n891}).
We compared these observations with the less deep observations obtained as part of the Guaranteed Time Key Project Very
Nearby Galaxy Survey (VNGS; KPGT\_cwilso01\_1; PI: C. D. Wilson) finding agreement between the two datasets.
Also, deeper observations are available from the open-time program OT1\_sveilleu\_2 (PI: S. Veilleux).
However, the field of view is too narrow and border effects would be introduced in the vertical profile; we therefore decided not to use this dataset.
% Actually we think that the background subtraction at the time of data reduction is 
% not well down because the map is too narrow!
%pa = 256
%rot 90 - 22.9
%proposal = OT2_emurph01_3

{\it Herschel} SPIRE (\citealt{2010A&A...518L...3G}) photometric observations at 250 \mic\,\,were
obtained as part of the VNGS (Obs ID: 1342189430).
The galaxy was observed in large map mode, covering an area of $20'$x$20'$ centred on the
object with two cross-scans, using a $30$ s$^{-1}$ scan rate. Data
were reduced with HIPE using the SPIRE
calibration tree SPIRE\_CAL\_12\_3 and the standard pipeline destriper to remove baselines. 
The map was then 
produced using the na{\"i}ve mapmaking procedure within HIPE. The resulting image
has a uniform background and pixel size of $6''$ (Fig.~\ref{fig:n891}).
%PA = $22.9$ (as estimated by PA fitting performed by \citealt{2011A&A...531L..11B}
%and \citealt{2014A&A...565A...4H}) 
In-beam flux densities are converted to surface brightness assuming an effective beam solid
angle $\Omega = 465$\,arcsec$^2$ (SPIRE Handbook, 2014\footnote{\url{http://herschel.esac.esa.int/Docs/SPIRE/pdf/spire_om.pdf}}).
SPIRE observations at 350 and 500 $\mu$m were also obtained as part of the VNGS but not
used here since the instrument resolution at these wavelengths is too low for our purposes.

{\it Spitzer} IRAC individual Basic Calibrated Data (BCD) frames were taken from the combined
observations (8 Astronomical Observing Requests) acquired by G. Fazio ({\it Spitzer}
Cycle 1, PID 3), and  processed with version 18.25.0 of the SSC pipeline. 
The data were acquired in High Dynamic Range (HDR) mode, so that each frame
consisted of one short integration and two long integrations.
The 442 long-integration time BCDs were combined into a single mosaic with
MOPEX (\citealt{2005PASP..117.1113M}). 
Bad pixels, i.e. pixels masked
in the data collection event (DCE) status files and in the static masks
(pmasks), were ignored. Additional inconsistent pixels were removed by means of
the MOPEX outlier rejection algorithms. We relied on both the box and the dual outlier
techniques, together with the multiframe reject algorithm. The frames were corrected
for geometrical distortion and projected on to a north-east coordinate system
with pixel sizes of $0\farcs6$, roughly half the size of the original pixels. 
The final mosaics were obtained with standard linear interpolation. The same was done
for the uncertainty images, i.e. the maps for the standard deviations of the
data frames. 
After experimenting with various HDR correction algorithms, 
we neglected this correction as it resulted in inferior final mosaics.
Despite several image reduction attempts,
the IRAC Channel 4 image (see Fig. \ref{fig:n891}) still shows a residual
column bleed, visible toward the southwest part of the image.
However, these artifacts do not affect our analysis as they are masked out when necessary.
In order to obtain the ``dust map'' at 8 $\mu$m we subtract the stellar component
following the prescription by \cite{2004ApJS..154..253H}:
we scale the IRAC 3.6 $\mu$m map by 0.232 and we subtract it from the observed 
IRAC 8 $\mu$m map on a pixel-by-pixel basis.

{\it Spitzer} MIPS data at 24 \mic\,\,were processed by Bendo et al. (2012)
and the final image has a pixel size of $1\farcs5$ (Fig.~\ref{fig:n891}).

PACS and SPIRE images were created on a grid rotated of an angle equal to the galaxy's 
position angle (PA = $22.9$, as estimated by PA fitting performed by \citealt{2011A&A...531L..11B}
and \citealt{2014A&A...565A...4H}) in order to have the galactic disc on the horizontal
direction.
IRAC and MIPS data were rotated by the same angle in post-processing after data reduction.
For each image, the mean sky flux estimated from different background apertures was
subtracted.

\section{Dust emission profile}
\label{sect:dust_em}

In order to test the possibility to quantify emission both from a thin dusty disc and from a 
thicker halo dust component we consider a general vertical profile of the form:
\begin{equation}
\label{eq:vprof}
I_{\rm d} (z) = I_{\rm d, 1} \exp \left(-\frac{z}{z_{\rm d,1}}\right) + I_{\rm d ,2} \exp \left(-\frac{z}{z_{\rm d,2}}\right),
\end{equation}
where $I_{\rm d, 1} + I_{\rm d, 2}$ is the dust surface brightness at the galactic midplane,
and $z_{\rm d,1}$ and $z_{\rm d,2}$ are the scaleheights of two dust components.
We then convolved this general vertical profile with the PSF of each instrument used in our analysis
and compared the resulting profiles with observations.

\subsection{Point and ``Line'' Spread Functions}

For an extended object, the observed surface brightness distribution, $g(i,j)$,
is the result of the convolution of an intrinsic surface brightness distribution,
$f(i,j)$, with the instrument ${\rm PSF}(i,j)$,
\begin{align}
\label{eq:conv1}
g(i,j) & = \left[ {\rm PSF} \Conv f \right](i,j)\nonumber\\
 & = \sum_{s = -N_i/2}^{N_i/2} \sum_{t = -N_j/2}^{N_j/2} f(s,t) \,{\rm PSF}(i-s,j-t),
%
%g(x,y) = \iint du\,dv\, f(u,v) \; {\rm PSF}(x-u,y-v).
\end{align}
where $i$ and $j$ are two perpendicular directions and $N_i$ and $N_j$ are the number 
of pixels along these two directions, respectively.
A (resolved) edge-on galaxy is a particular case of extended emission source where 
the surface brightness has a smooth gradient along the galactic disc 
compared to
the perpendicular direction (here defined as $i$  and $j$ directions, respectively).
Assuming that the intrinsic emission is constant along $i$ and infinitely extended
(or at least much more extended than the considered region), so that 
$f(i,j)=f(\overline{\imath},j)=f(j)$, with $\overline{\imath}$ any position along $i$, we can 
write:
\begin{align}
\label{eq:conv2}
g(\overline{\imath},j) & = \sum_{t = -N_j/2}^{N_j/2} f(t) \left[ \sum_{s = -N_i/2}^{N_i/2} {\rm PSF}(\overline{\imath}-s,j-t) \right]\nonumber\\
			        & = \sum_{t = -N_j/2}^{N_j/2} f(t)\, {\rm LSF}(j-t)
%g(\overline{x},y) &= \int dv\, f(v) \, \int du\; {\rm PSF}(\overline{x}-u,y-v) \nonumber \\
%       &= \int dv\; f(v) \; {\rm LSF}(y-v) 
\end{align}
where the {\em ``Line" Spread Function} (LSF)
\[
{\rm LSF}(j) = \sum_{i = -\infty}^{\infty} {\rm PSF}(i,j)
\]
represents a vertical cut of the two-dimensional image resulting from
the convolution of the instrument PSF with a infinitely thin line.
Under the above assumptions (see Appendix~\ref{app:model}
for a test on the assumed geometrical model), a vertical profile extracted from the image
at a position $\overline{\imath}$ can be compared equivalently to
\begin{itemize}
\item a profile, $g(\overline{\imath},j)$, extracted from a modelled image obtained
from the convolution of the two-dimensional intrinsic surface brightness distribution,
 $f(i,j)$, with the PSF (Eq.~\ref{eq:conv1});
\item a one-dimensional model, $f(j)$, convolved with the LSF (Eq.~\ref{eq:conv2}).
\end{itemize}
The first approach was used by \citet{2011A&A...531L..11B}, while the second
by \citet{2013A&A...556A..54V} and \citet{2014A&A...565A...4H} (though without
a formal demonstration of its validity). In this work we use the latter.

\subsection{{\it Spitzer} and {\it Herschel} PSFs}

\begin{figure*}[ht]
\begin{center}
%this was done using convolution/compare_profiles
\includegraphics[trim=0mm 5mm 0mm 0mm,scale = 2.]{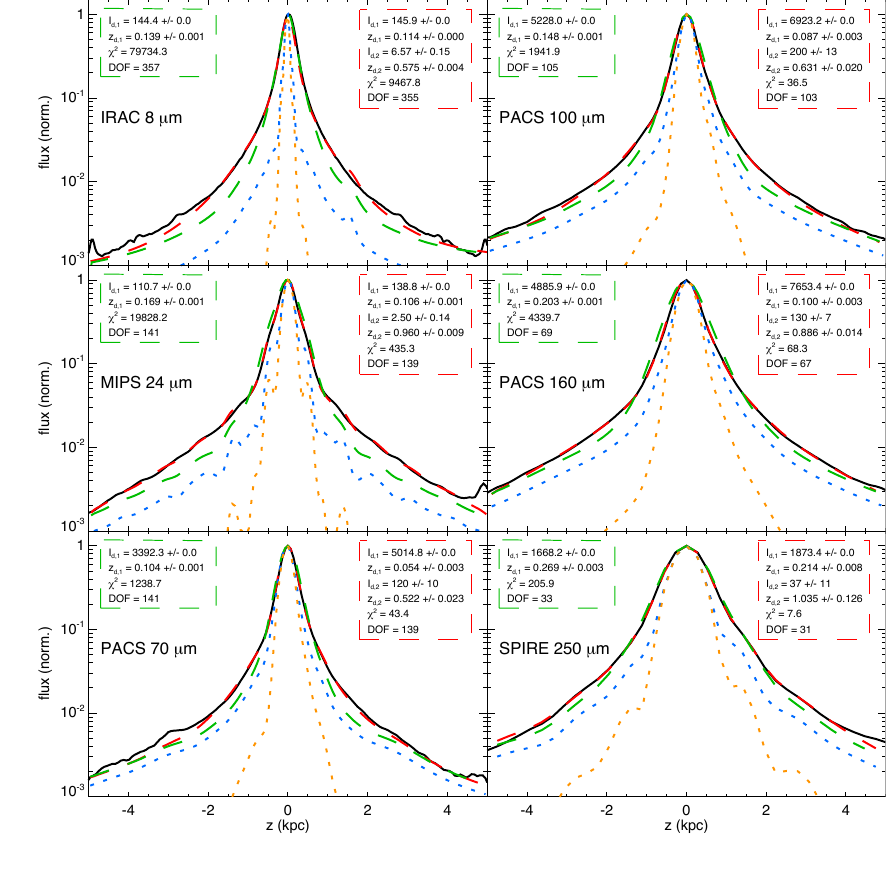}
\caption{Normalised vertical profiles for observations (black solid lines). The green dashed lines are the result of the 
convolution of the one-component dust emission profile with the instrument LSF while the red dashed 
lines are the result of convolution of the two-component dust emission profile. 
Normalisation values are 
82.03, 48.17, 1119.47, 2023.22, 1798.00, 
630.58 MJy/sr for IRAC 8, MIPS 24, PACS 70, 100, 
160 and SPIRE 250 $\mu$m observed profiles, respectively.
%Normalisation values are 
%82.03 (152.47), 48.17 (141.26), 1119.47 (5135.18), 2023.22 (7124.01), 1798.00 (7783.55), 
%630.58 (1910.85) MJy/sr for IRAC 8, MIPS 24, PACS 70, 100, 
%160 and SPIRE 250 $\mu$m observed (intrinsic) profiles, respectively.
Fit results are indicated in the green and red boxes,
respectively (scaleheights expressed in kpc and the intrinsic surface brightness in MJy/sr). 
LSF and PSF vertical profiles are illustrated in blue and orange dotted lines, respectively.}
\label{fig:vprofile_comb}
\end{center}
\end{figure*}

Knowing the PSF of an instrument with a good precision is key 
to perform convolution (or deconvolution) operations on images and we therefore 
carefully consider several details to have the best estimate of each PSF.
In particular, the vertical profile of this source is resolved but its extent is comparable to
the FWHM of the PSF, therefore making the choice of the right set of PSFs a critical issue.

Theoretical {\it Herschel} PACS PSFs are released by the PACS Instrument Control Centre (ICC)
assuming a spectral slope $\alpha = -4$ in $F_{\lambda}$, which corresponds to the ideal PSF
that would have a point source with a very high temperature that emits with a blackbody spectrum, $B_{\nu}(T)$ 
(we will refer to these PSFs as ``A(std)'').
However, dust emits with a modified blackbody spectrum, 
$\left( \nu \right)^{\beta} B_{\nu}(T)$,
%\begin{equation}
%\label{eq:modBB}
%I_{\nu} = \tau_{\nu_0} \left( \nu / \nu_0 \right)^{\beta} B_{\nu}(T)
%\end{equation}
where $\beta$ is the spectral
index and $B_{\nu}(T)$ is a blackbody radiation for an equilibrium temperature $T$.
Dust in galaxies is expected to be in thermal equilibrium at an 
average temperature $T \sim 20$\,K and a spectral index 
$\beta \sim 2$, leading to a modification of the effective PSF. 
We therefore modified the A(std) PSFs according to the dust spectrum assuming $T = 20$\,K and $\beta = 2$ 
(see Appendix~\ref{app:PSF_mod} for a detailed description of the method used).
We will refer to these PSFs as ``A(mod)''.

We compared both the standard and modified PACS PSFs with observations of Vesta and Mars 
(point-like sources in PACS bands)  
and we noted relevant differences in the emission profile.
For this reason, we combined these observations to have a better estimate of the observed 
PSF (``B(std)'', see Appendix~\ref{app:psfobs} for details).
As for the theoretical PSFs from the ICC we modified the observed PSFs according to the 
dust spectrum assuming the same temperature and $\beta$ values (``B(mod)'').

%Herschel PACS psfs 
%- ICC, standard and modified according to...
%- observed, standard and modified according to...
%PACS PSFs from the ICC are based on models and narrower than observations...
%In order to better estimate them we used observations of Vesta and Mars combined together.
%Also, the dust spectrum is different from the spectrum of Mars or Venus, in particular, its temperature is 
%lower and beta higher.
%We modified the PSFs according to the spectrum....

{\it Herschel} SPIRE PSFs are released by the ICC and are obtained from scan-map data of Neptune and represent therefore empirical beam maps, ``A(std)''.
Following the same method as for the PACS PSFs, we modified them according to the dust spectrum assuming
the same temperature and $\beta$ values (see Appendix~\ref{app:PSF_mod} for details, ``A(mod)'').

The IRAC Instrument Support Team noted that IRAC PSFs are undersampled and then developed 
Point Response Functions (PRFs) which combine the PSF information and the detector sampling.
The PRFs were generated from models refined with in-flight calibration test data involving a 
bright calibration star observed at several epochs, making them good estimates of the observed beam\footnote{IRAC PRFs are available at
\url{http://irsa.ipac.caltech.edu/data/SPITZER/docs/irac/calibrationfiles/psfprf}}.

Theoretical PSFs for the {\it Spitzer} MIPS camera were generated by \cite{2011PASP..123.1218A} using the software 
STinyTim\footnote{MIPS PSF FITS files are available at \url{http://www.astro.princeton.edu/~ganiano/Kernels.html}}.
They assumed a 
blackbody source at $T = 25$\,K and used a pixel 
size of $0.5''$ (we will refer to the MIPS 24 \mic\,PSF as ``A(std)'').
As the optics are very smooth, the PSFs generated in this way matches the observed PSFs
with a sufficient precision (MIPS Instrument Handbook).

After several tests, we define our best estimate of PSFs as A(std) for IRAC and MIPS, 
B(mod) for PACS and A(mod) for SPIRE.
All these PSFs cover a large dynamic range ($\gtrsim10^6$) and are defined until large radii from the centre ($>3$\,arcmin); 
thus they should not miss any faint point-source extended emission.

None of the PSFs described above have been circularised and the relative orientation 
between PSF and image needs to be taken into account.
This is achieved by rotating the PSF
image so that the direction of the spacecraft Z-axis aligns with the direction of the spacecraft
Z-axis on NGC 891 images (indicated by the white arrow for all images in Fig.~\ref{fig:n891}).
Subsequently, to obtain the LSF, for each vertical distance, pixels are summed along 
the direction of the galactic major axis.
The LSF is always broader than the PSF because of the integrated contribution
of the PSF Airy rings off the main beam (PSF and LSF would have the same FWHM, 
instead, if the PSF were Gaussian and circular). 
However, some positions along the major axis of an edge-on galaxy might
be dominated by the emission of unresolved sources (such as bright
star-forming regions), rather than from a more diffuse disc. The vertical 
profile above those regions would then appear narrower. 
We will discuss the impact of this issue on our analysis in Section~\ref{sect:vert_var}.
A vertical cut of LSFs and PSFs for all the considered instruments are shown in Fig.~\ref{fig:vprofile_comb} in
blue and orange dotted lines, respectively.

\subsection{Vertical profile extraction and error estimation}
\label{sect:vertprof}
In order to extract vertical profiles from observations we consider strips of (-7,+7) kpc 
parallel to the major axis of the galaxy 
(corresponding to $-2.5'$,$+2.5'$) from the centre of the galaxy ($\alpha$ =
2$^{\rm h}$22$^{\rm m}$33$\fs$0, $\delta$ = 42$^{\circ}$20$'$57$\farcs$2, 
J2000.0, blue cross in Fig.~\ref{fig:n891}),
where the radial profile along the midplane is relatively constant (emission always larger 
than $\sim 1/10$ of the peak, see Fig.~\ref{fig:radial_multi}) in all bands considered.
For each distance from the midplane (the vertical step is one pixel) we then calculate the median 
emission over all the pixels in the row within the specified area.
Vertical profiles are normalised to unity and shown in Fig.~\ref{fig:vprofile_comb} in black solid lines.

For a comparison of the convolved dust emission profile with observations a
careful estimation of the uncertainties is needed.
Following \cite{2012A&A...543A.161C}, \cite{2013PASP..125.1126R} and \cite{2014MNRAS.440..942C} we identified 
two main sources of uncertainties.
The first source is the instrumental noise, $\sigma_{\rm instr}$, based on the error map. It depends on the number of scans
crossing a pixel and can be estimated using the sum
in quadrature of the error map pixels in the measurement region.
The second source is from background noise, $\sigma_{\rm sky}$, and can be estimated as:
\begin{equation}
\sigma_{\rm sky}^2 = N_{\rm ap}\sigma_{\rm skypix}^2+N_{\rm ap}^2\sigma_{\rm skymean}^2,
\end{equation}
where $\sigma_{\rm skypix}$ is the RMS of the fluxes in the chosen aperture\footnote{$\sigma_{\rm skypix}$ defined in this way 
includes not only the RMS of the background sky but also possible variations of the flux due to the non-flatness of the radial profile of the galaxy at the different distances from the midplane.}, 
$N_{\rm ap}$ is the number of pixels in the aperture and $\sigma_{\rm skymean}$ is the standard deviation of the mean value of the sky
measured in different apertures far from the galaxy.
For each vertical distance, the aperture is defined as the row of pixels over which we calculate the median emission
(in this case from -7 kpc to +7 kpc from the galactic centre).

We then obtain the total uncertainty, $\sigma_{\rm tot}$, as:
\begin{equation}
\sigma_{\rm tot} = \sqrt{\sigma_{\rm instr}^2+\sigma_{\rm sky}^2}.
\end{equation}
It has to be noted that, since we are not directly comparing images at different wavelengths, calibration errors are
not considered here but will be later in Section~\ref{sect:dust_temp}.

%We consider a general vertical profile of the form:
%\begin{equation}
%\label{eq:vprof}
%L_{\rm d} (z) = L_{\rm d, 1} \exp \left(-\frac{z}{z_{\rm d,1}}\right) + L_{\rm d ,2} \exp \left(-\frac{z}{z_{\rm d,2}}\right),
%\end{equation}
%where $L_{\rm d, 1} + L_{\rm d, 2}$ is the dust luminosity at the galactic midplane,
%and $z_{\rm d,1/2}$ are the scaleheight of two dust components.
%We then convolved this general vertical profile with the PSF of each instrument used in our analysis
%and compared the resulting profiles with observations.
%We consider two possibility, a single dust component (1c, $L_{\rm d ,2} = 0$) or two separated dust components (2c).
%The resulting intrinsic and convolved vertical profiles for PACS 100 \mic\,are shown in Fig.~\ref{fig:vprofile_PACS100}.
%Intrinsic vertical profiles are shown in solid lines (yellow and red for 1c and 2c, respectively) while convolved vertical profiles
%are represented by dotted lines (with the same color coding).
%Fitting parameters are indicated for the two cases, 
%dust luminosities at the galactic midplane are normalised to unity and scale 
%heights are expressed in kpc.

\subsection{Fitting results}
\label{sect:fitting}

We consider two possibilities, a single dust component (1c), for which $I_{\rm d ,2} = 0$ in Eq.~\ref{eq:vprof} or 
two separated dust components (2c).
We then convolve the intrinsic vertical profile to the corresponding instrument LSF (we use our best set of PSFs here) 
and used a $\chi^2$-minimisation routine to fit the observed vertical profiles.
The resulting convolved vertical profiles are shown in Fig.~\ref{fig:vprofile_comb}
in green (1c) and red (2c) dashed lines, normalised to unity at the galactic midplane.
%Intrinsic vertical profiles are shown in solid lines (orange and red for 1c and 2c, respectively) while convolved vertical profiles
%are represented by dotted lines (with the same color coding).
Fitting parameters are indicated for the two 
cases\footnote{\cite{2014A&A...565A...4H} also performed single components fits at 100, 160 and 250 $\mu$m,
finding vertical scalelengths larger by up to a few times those shown in Fig.~\ref{fig:vprofile_comb} of
this work. However, an inspection of the vertical profiles presented in Fig.~\ref{fig:vprofile_comb} of
{\em their} paper, and in particular of their LSF, suggests that the vertical scales
have been accidentally multiplied by constant factors. If the factors were removed,
their scalelengths would be 0.14, 0.15 and 0.23 kpc at 100, 160 and 250 $\mu$m,
respectively. The values are in much better agreement with our estimates, if we
consider also the different treatment of the PSF we do in this work.}, the intrinsic dust surface brightness is
expressed in MJy/sr and scaleheights in kpc.
Both visually and from fitting parameters, we note that for all considered instruments 
the two dust component case best fit the observations.
In particular, $\chi^2_{\rm red} = \chi^2/ {\rm DOF}$ 
for the two dust component case is $\lesssim 5$ (with the exception of IRAC 8 
$\mu$m data where $\chi^2_{\rm red} \sim 30$), while
for the case of a single dust component is always $\gtrsim 5$ times larger.
%An exception is made for IRAC 8 $\mu$m data where $\chi^2_{\rm red} \sim 30$ and $\sim 220$ 
%for two- and one-component cases, respectively. 
Using different sets of PSFs, the assumption of two dust components always leads to
a better fit to observations for all the considered wavelengths.

%\begin{table}
%\caption{$\chi^2_{\rm red}$ values for one (1c) or two (2c) dust components.}             
%\label{table:oneparam}      
%\centering          
%\begin{tabular}{ l c c c c c }
%\hline\hline
%& MIPS24 & PACS70 & PACS100 &PACS160 & SPIRE250\\
%\hline
%1c &  140.6&8.8 & 18 & 63 & 6.2\\
%2c & 3.14 & 0.31 & 0.35 &1.02 &0.25 \\ 
%\hline
%\end{tabular}
%\end{table}

%\subsection{Fitting results: scaleheights}

%Here we assume two dust components.
%We go though all the different choices of PSfs and we say which are the best.
%And we say that even assuming other PSFs the Xred is always much better for two components. It does not
%affect results!

\begin{table*}[t!]
\caption{Dust emission scaleheights (for thin and thick discs) as measured using IRAC, MIPS, PACS and SPIRE instruments. 
The PSFs used for the convolution are indicated by A/B (std, obs), see text for details.\vspace{.2cm}}
% THIS IS UPDATED!             
\label{table:scaleheights}      
\centering          
\begin{tabular}{@{\extracolsep{1mm}} l c c c c c c }
\hline\hline
& \multicolumn{2}{c}{IRAC 8 \mic} & \multicolumn{2}{c}{MIPS 24 \mic} & \multicolumn{2}{c}{PACS 70 \mic}  \\
\cline{2-3}
\cline{4-5}
\cline{6-7}
& $z_{\rm d,1}$ & $z_{\rm d,2}$& $z_{\rm d,1}$ & $z_{\rm d,2}$ & $z_{\rm d,1}$ & $z_{\rm d,2}$ \\
& (kpc) & (kpc)& (kpc) & (kpc) & (kpc) & (kpc) \\
\hline
%ICC, std chi2,red  =  3.13  0.56  0.83  4.7  0.25
%ICC, mod chi2,red  =  -  0.54  0.797  4.486  0.245 
%obs, std chi2,red  =   - 0.23 0.38  1.64 -
%obs, mod chi2,red  =  -  0.31 0.354 1.0189 -
A(std) &$0.114\pm 0.01$ & $0.575\pm 0.001$ & $0.106\pm0.001 $ & $0.96\pm0.01 $ & $0.119 \pm 0.002 $ & $0.76\pm0.03$  \\
A(mod)& - & -  & - & - & $0.114\pm0.002$  & $0.76\pm0.03$  \\
\hline
B(std) & - & - & - & - & $0.074\pm0.002$  & $0.56\pm0.02$    \\
B(mod)& - & -  & - & - & $0.054\pm 0.003$ & $0.52\pm0.02$  \\
\hline
{\color[rgb]{1,0,0} Region X} &$0.145\pm0.01$  &$ 0.822\pm0.01$& $0.140\pm0.001 $ & $1.36\pm0.01 $ & $0.085 \pm 0.007 $ & $1.28\pm0.32$  \\
\hline\\
\end{tabular}
\begin{tabular}{@{\extracolsep{1mm}} l c c c c c c }
\hline\hline
&  \multicolumn{2}{c}{PACS 100 \mic} & \multicolumn{2}{c}{PACS 160 \mic} & \multicolumn{2}{c}{SPIRE 250 \mic}  \\
\cline{2-3}
\cline{4-5}
\cline{6-7}
 & $z_{\rm d,1}$ & $z_{\rm d,2}$& $z_{\rm d,1}$ & $z_{\rm d,2}$ & $z_{\rm d,1}$ & $z_{\rm d,2}$ \\
& (kpc) & (kpc)& (kpc) & (kpc) & (kpc) & (kpc) \\
\hline
A(std) & $0.138\pm 0.002$ & $0.88\pm0.03$ & $0.191\pm0.002$ & $1.17\pm0.02$  & $0.218\pm0.008$ & $1.10\pm0.13$\\
A(mod)& $0.135\pm0.002$ & $0.90\pm0.03$ & $0.188\pm0.002$ & $1.21\pm0.02$ & $0.214\pm0.008$ & $1.03\pm0.13$  \\
\hline
B(std) & $0.101\pm0.002$ & $0.67\pm0.02$& $0.115\pm0.003 $ & $0.91\pm0.01$ & -  & -\\
B(mod) & $0.087\pm0.003$ & $0.63\pm0.02$ & $0.100\pm0.003$ & $0.89\pm0.01$ & - & -  \\
\hline
{\color[rgb]{1,0,0} Region X} & $0.114\pm 0.004$ & $1.02\pm0.10$ & $0.140\pm0.003$ & $1.14\pm0.03$ & $0.239\pm 0.007$ & $1.40\pm 0.12$  \\
\hline
\end{tabular}
\end{table*}

Hereafter we therefore consider only the case of two dust components.
We have convolved the intrinsic vertical profile with all the PSFs described earlier
and we report the scaleheights of the thin and thick components for different
PSFs and for all the considered instruments in Table~\ref{table:scaleheights}.
% illustrates the scaleheights of the thin and thick components for different
%PSFs and for all the considered instruments.
The errors indicated in this table result from the $\chi^2$-minimisation technique.
Region X refers to a particular region in NGC 891, at a projected radial distance $h_{\rm X} \sim 5.6$\,kpc 
SSW from the galactic centre, where the influence from unresolved sources was estimated to be lower
(see Section~\ref{sect:vert_var}). Scaleheights for this region are 
illustrated only for our best estimates of the PSFs.

While the use of standard (std) or modified (mod) PSFs does not affect the scaleheights,
a substantial change is achieved using A or B PSFs for PACS (see Appendix~\ref{app:plots_scaleheights} for more details).
Moreover, scaleheights for region X are always larger (by more than a factor two in some cases) 
than estimates integrated over the global ($\pm 7$\,kpc) strip described in Sect.~\ref{sect:vertprof}.

\begin{figure}[h!]
\begin{center}
\includegraphics[scale=1]{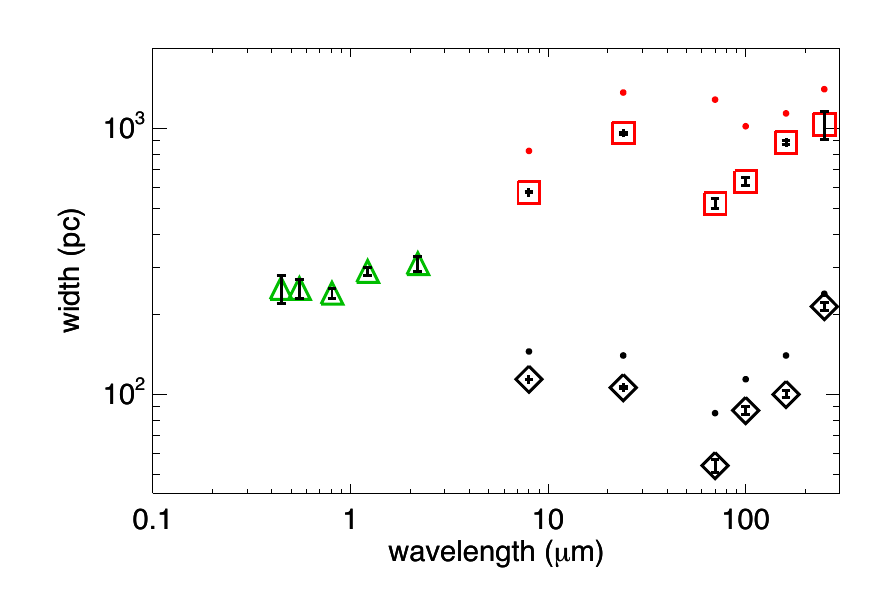}
%\includegraphics[scale=1]{images/widths_ext_noncirc_nol_modpsf_comb.ps}
%widths_ext_noncirc_nol_modpsf_comb_z2.ps
\caption{Scaleheights of the thin (black diamonds) and thick (red squares) dust components. 
Green triangles represent the dust scaleheights as estimated by \cite{1999A&A...344..868X}.
Black and red dots are the scaleheights as measured in region X (see Section~\ref{sect:vert_var}).
Our best set of PSFs is used for convolution.}
\label{fig:widths_nol}
\end{center}
\end{figure}

Fig.~\ref{fig:widths_nol} shows the two scaleheights (black and red symbols for thin and thick components, respectively)
for our best set of PSFs.
Error bars indicate the errors resulting from the $\chi^2$-minimisation technique.
Results from fitting optical/NIR images with radiative transfer models are shown in green 
triangles (\citealt{1999A&A...344..868X}).
Red and black dots are the dust scaleheights for region X.

Scaleheights obtained from observations at galactic scale show a wavelength gradient 
for $\lambda \geqslant 70\,\mu$m for both the thin and thick components.
However, estimates from region X show a smoother gradient for the scaleheight of the thin component 
and an almost constant scaleheight for the thick component with variations of $\lesssim 30\%$.
%variations of $\lesssim 30\%$ for the thick component scaleheight.
The observed gradient in the scaleheight of the thin component may indicate that 
(see discussion in Section~\ref{sect:discussion}):
\begin{enumerate}
\item the dust temperature quickly drops within its scaleheight leading to the dimming of the FIR emission or
\item also region X is affected by the emission of point-like sources that emit at relatively
warm temperatures and therefore tend to reduce the effective scaleheight depending on the relative
wavelength.
\end{enumerate}

Furthermore, the scaleheights of the thin component obtained from 8 and 24-$\mu$m emission are 
very similar both from measurements at galactic scale and in region X.
This indicates that these estimates are affected by the presence of point sources 
to the same degree.
On the other hand, the thick component at 24 $\mu$m is $\sim 1.6$ times more
extended than at 8 $\mu$m.

\section{Vertical profile variation}
\label{sect:vert_var}

\begin{figure}[here]
\begin{center}
%profile/plot_radial.pro
\includegraphics[scale=1]{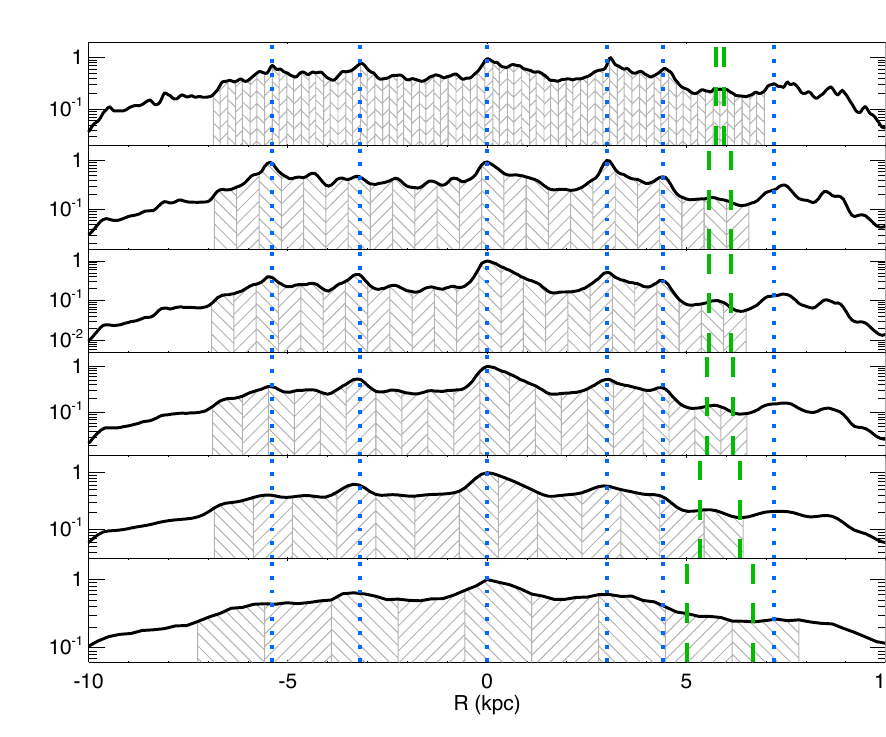}
\caption{Radial profiles for IRAC 8 \mic, MIPS 24 \mic, PACS 70, 100 and 160 \mic\,and SPIRE 250 \mic\,(from top to bottom).
The position of the main peaks in emission (blue dotted lines) and the region X (green dashed lines) are indicated.}
\label{fig:radial_multi}
\end{center}
\end{figure}

To inspect the variability of the vertical profile across the galaxy we subdivided the midplane into 
radial regions. Each region has been chosen to have a radial width that is twice as large as the FWHM of
the instrument PSF.
Radial profiles for all considered wavelengths are shown in Fig.~\ref{fig:radial_multi} and shaded regions indicate 
the radial portions into which the midplane has been divided.
The position of the main peaks in emission is highlighted by the blue dotted lines.

We define ``region X'' the vertical strip cutting the midplane at a projected distance $h_{\rm X}$ SSW
from the galaxy centre. This region has been chosen to have a radial width twice as
large as the FWHM of the instrument PSF and not to comprise
any of the main emission peaks for all the wavelengths considered.
Furthermore, the continuum-subtracted H$\alpha$ emission-line image of this galaxy
(Fig.~10 in \citealt{2000AJ....119..644H}) 
shows that most of the galactic disc is bright in H$\alpha$, indicating the presence of 
several large H{\sc ii} regions.
However, at the position corresponding to our region X, the H$\alpha$ emission is the lowest
observed in the midplane, therefore suggesting a low contamination by 
bright and hot dust emission from H{\sc ii} regions and making this region unique in this source.
In Fig.~\ref{fig:radial_multi} we indicate region X with green dashed lines.

\begin{figure}[here]
\begin{center}
\includegraphics[scale=1]{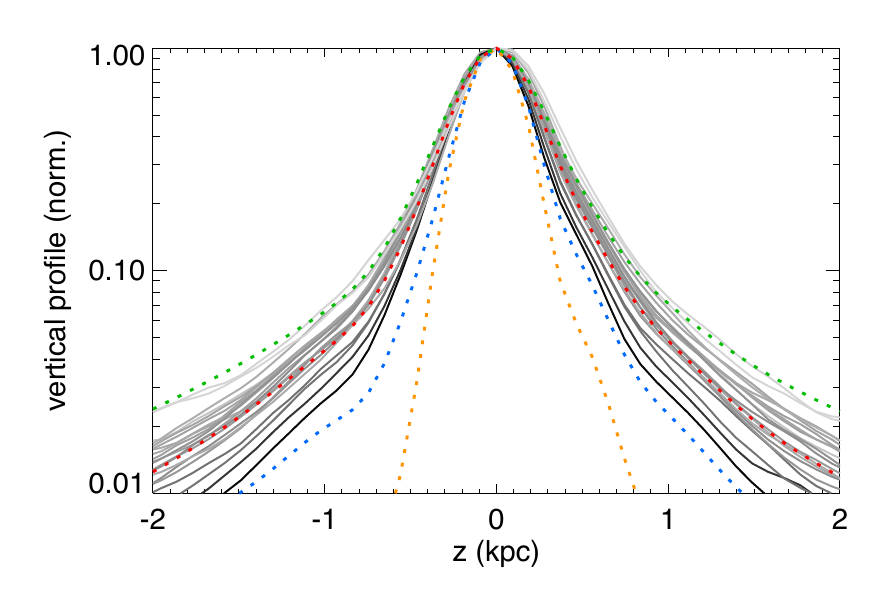}
\caption{Vertical profiles at 100 \mic\,as measured in different radial regions. A vertical cut of the PSF (orange) and of the LSF (blue) is indicated.
The vertical profile for the whole galaxy and for region X are shown in red and green, respectively.}
\label{fig:profiles_PACS100}
\end{center}
\end{figure}

We extract the vertical profile for each of the considered regions and for all the considered wavelengths.
In Fig.~\ref{fig:profiles_PACS100} we show the vertical profiles for PACS 100 $\mu$m as an example.
The intensity of the shades of grey indicates the surface brightness of dust emission (black being the brightest region).
A vertical cut of the PSF and of the LSF are indicated by orange and blue lines, respectively. 
The convolved vertical profile obtained from fitting the observed vertical profile of the whole galaxy 
(-7 kpc to 7 kpc from the centre)
is shown in red dotted line, while the green dashed line indicates the convolved vertical profile obtained from fitting 
observations in region X.

Our analysis shows that there is some variation
in the vertical profiles. Bright regions have a narrower
vertical profile indicating that these regions might be dominated 
by isolated point sources 
or by a very thin disc not resolved in the vertical profile 
%(showing an LSF profile; see Appendix A ...
%are dominated by the emission of point sources 
(see Appendix~\ref{app:model} for further details).
On the contrary, in region X, where the radial profile in the midplane is rather flat,
the extracted vertical profile is close to the widest observed vertical profile.
Region X is therefore affected to a low degree by the emission of unresolved sources and
for this reason the vertical profile extracted in this region represents a better estimate of the real extent of
extra-planar dust with respect to the vertical profile extracted at galactic scale.
Scaleheights of the thin and thick discs retrieved from fitting observations in region X 
are indicated in Fig.~\ref{fig:widths_nol} (see the black and red dots, respectively) and listed in 
Table~\ref{table:scaleheights}.

\section{Dust Spectral Energy Distributions}
\label{sect:dust_temp}

%I asked myself: can we do this for all regions along the midplane?
%The answer is no, because we cannot consider regions dominated by point sources, in these regions the
%vertical profile is a Dirac delta.

By fitting the median vertical profiles of region X for each wavelength,
we can construct dust Spectral Energy Distributions (SED) for each vertical distance
from the galaxy midplane.
We computed the uncertainties on the surface brightness in this way:
considering the parameters for region X listed in Table~\ref{table:scaleheights},
we randomly generated a large number of profiles following a Gaussian 
distribution centred at $z_{\rm d,1}$ and $z_{\rm d,2}$ and with the standard deviation
given by the error on parameters.
For each vertical distance we then computed the mean intensity and the standard deviation of the distribution which
will represent the error on the mean intensity.
Furthermore, since we are comparing fluxes in different bands, 
we need to include the calibration errors, $\sigma_{\rm cal}$,
which will be added in quadrature to the errors computed earlier.
We consider a calibration uncertainty of 3\% (IRAC Instrument Handbook), 4\% (\citealt{2012MNRAS.423..197B}), 
7\% (\citealt{2014ApJ...789L..38B}) and 7\% (SPIRE Handbook Version 2.5) for IRAC, 
MIPS, PACS and SPIRE data, respectively.

\subsection{Dust modelling}
\label{sect:dust_modelling}

For each vertical distance we fit the intrinsic spectrum from 8 $\mu$m to 250 $\mu$m
using the Jones et al. (2013) dust 
model and the DustEM code\footnote{\url{http://www.ias.u-psud.fr/DUSTEM/}} to compute the dust SED.
During this procedure we use the standard \cite{1983A&A...128..212M} interstellar radiation field and let 
the following parameters vary:
\begin{itemize}
\item $G_0$, a parameter indicating the intensity of the radiation field,
\item $a_{\rm min}$, the minimum size of small grains in the model,
\item $m_{\rm SG} / m_{\rm LG}$, the ratio between the mass of small grains and that of large grains.
\end{itemize}
In this way we can constrain the radiation field intensity and the fraction of small grains.

Furthermore, in order to have information on the average dust temperature and on its opacity, 
we fit the SED from 70 $\mu$m to 250 $\mu$m to a single-temperature modified blackbody radiation (MBB):
\begin{equation}
\label{eq:modBB}
I_{\nu} = \tau_{\nu_0} (\nu/\nu_0)^{\beta} B_{\nu}(T),
\end{equation}
where $\tau_{\nu_0}$ is the optical depth at a frequency $\nu_0$ corresponding to 100 \mic, 
$\beta$ is the spectral index at $\nu_0$ and $B_{\nu}(T)$ is a blackbody radiation for 
an equilibrium temperature $T$.

\begin{figure}[here]
\begin{center}
%image generated by convolution/conv_prof
\includegraphics[scale=1]{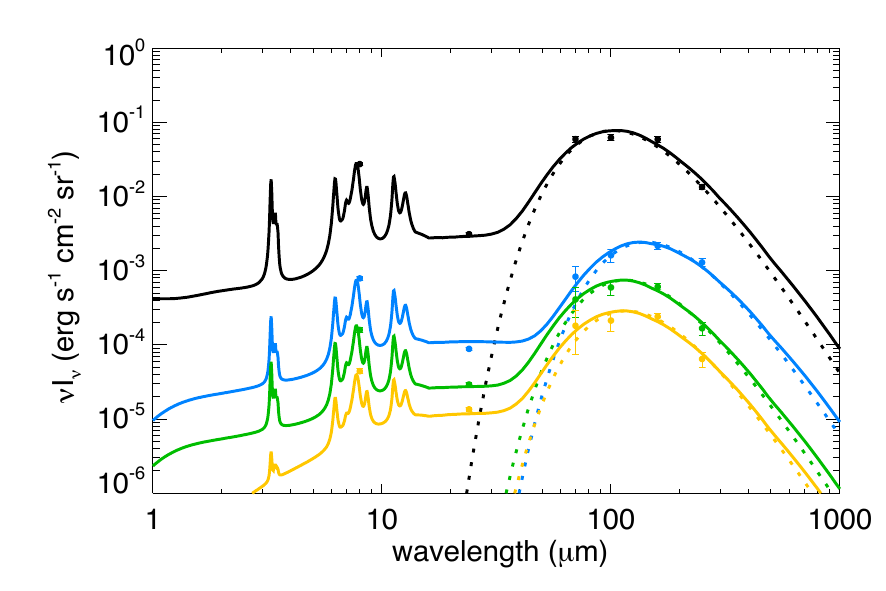}
\caption{Spectra for z = 0, 0.6, 1.5, 2.5 kpc (top to bottom). The fit using the Jones et al. (2013) dust 
model is indicated 
in solid lines. Spectra from 70 \mic\,to 250 \mic\, are fit with modified blackbody radiation 
with fixed $\beta = 2$ (dotted lines).}
\label{fig:spectra_multi}
\end{center}
\end{figure}

Fig.~\ref{fig:spectra_multi} shows the spectra at vertical distances $z = 0, 0.6, 1.5, 2.5$\,kpc (from top to bottom).
Solid lines represent the dust SED obtained from fitting the dust model while 
dotted lines represent the fit to a modified blackbody with fixed $\beta = 2$.
The emission intensity decreases with increasing $z$ and the modified blackbody peak position does not follow a clear trend.

Furthermore, we find that the value of $a_{\rm min}$ goes from 
$a_{\rm min} = (0.40\pm 0.05)$\,nm
at the galactic disc to $a_{\rm min} = (0.70 \pm 0.05)$\,nm at 2.5 kpc above the disc.
Consequently, the mass of small grains is reduced from $m_{\rm SG} / m_{\rm LG} = 
(6.2\pm 0.1)\times 10^{-2}$ to $m_{\rm SG} / m_{\rm LG} = (3.9\pm 0.1)\times 10^{-2}$.
This indicates that small grains are partly destroyed when moving from the galactic disc 
up to 2.5 kpc above it.
This could suggest that dust is entrained by interstellar shocks or galactic fountains
and during coupling to the gas small grains are partly eroded because of the high relative velocity 
with the gas.
Similar results have been observed in supernova remnants in the Large Magellanic Cloud 
(\citealt{2006ApJ...642L.141B,2006ApJ...652L..33W}) and obtained from
theoretical modelling (\citealt{2010A&A...510A..36M,2014A&A...570A..32B}).

As a further analysis, we include dust collisional heating due to the presence of the
hot halo gas, with an updated version of the DustEM code as described by \cite{2013A&A...556A...6B}.
From deep Chandra data, \cite{2013ApJ...762...12H} found hot halo gas at a temperature 
$T_{\rm gas} \sim 0.2$\,keV following a vertical exponential distribution:
\begin{equation}
n(z) = n_0 \exp(-z/z_{\rm hg})
\end{equation}
where $n_0 \sim 6\times 10^{-3}$\,cm$^{-3}$ and $z_{\rm hg} = 5\pm 2$\,kpc.
Large grains are not affected by this extra source of heating and their temperature 
is completely determined by the radiation field intensity.
On the other hand, small grains reach high temperatures and will emit more than 
if heated only by the radiation field.
This leads to $a_{\rm min} = (0.90 \pm 0.05)$\,nm and $m_{\rm SG} / m_{\rm LG} = 
(3.3\pm 0.1)\times 10^{-2}$ at 2.5 kpc above the disc, supporting the idea
that small grains are partly eroded moving from the galactic plane to 2.5 kpc above it.

\subsection{Dust temperature estimation}
\label{sect:dust_temp_est}

\begin{figure*}[ht]
\begin{center}
% This plot has been created using convolution/conv_prof.pro
\includegraphics[trim=10mm 0mm 0mm 0mm,scale = 1.1]{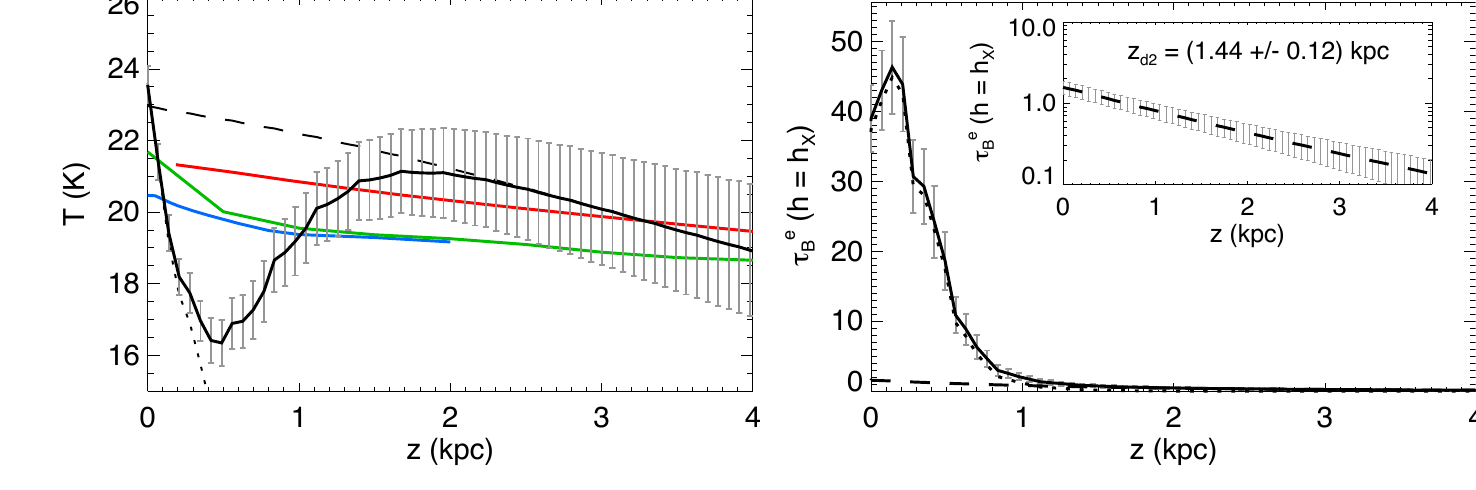}
\caption{Temperature (left panel) and $\tau_{\rm B}^{e} (h = h_{\rm X})$ (right panel) profiles
as obtained from MBB fitting (black lines).
Contributions from the thin (dotted lines) and thick
(dashed lines) discs are shown. Temperature profiles obtained using the \cite{2013A&A...558A..62J} and 
the radiative transfer models RT1, RT2 and RT3 are indicated in
blue, red and green lines, respectively. In the inset we show $\tau_{\rm B}^{e} (h = h_{\rm X})$ 
for the thick component and indicate its scaleheight.}
\label{fig:temp_profile_Pop}
\end{center}
\end{figure*}

One of the main parameters that we obtain from fitting data to a MBB is the luminosity-weighted
mean dust temperature at each
vertical distance which is shown in the left panel of Fig.~\ref{fig:temp_profile_Pop} in black solid line.
We also show the dust temperature for the thin and thick disc separately in dotted and dashed lines, respectively.
The dust temperature in the thin disc drops off rapidly while the dust temperature in the thick disc stays 
more elevated further from the midplane.
However, the dust temperature estimation depends strongly on the obtained
scaleheights for the two dust components. In particular, the steep gradient in the scaleheight of the thin component
might be affected by emission from a thin, unresolved, disc
%by the presence of point source emission in the midplane 
and this effect would be reflected on
the temperature estimation for $z \lesssim 1$\,kpc (see discussion in Sect.~\ref{sect:thincomp}).

As a comparison, exploiting the results from radiative transfer models and assuming a dust
model it is possible to calculate the resulting dust temperature.
We assumed the dust optical properties and size distributions from \cite{2013A&A...558A..62J} and 
we use different radiation fields, specific for this galaxy, obtained from radiative transfer models from the literature:
\begin{itemize}
\item \cite{2011A&A...527A.109P,2013MNRAS.436.1302P}, RT1
\item \cite{2008A&A...490..461B,2011A&A...531L..11B}, RT2
\item \cite{2012A&A...545A.124B}, RT3
\end{itemize}
Using the DustEM code we then compute the dust SED and fitting it to a MBB for $\lambda \gtrsim 70\,\mu$m
we estimate the average temperature of the large grains.

The resulting dust temperature profiles are shown in Fig.~\ref{fig:temp_profile_Pop} for  
RT1 (blue line), RT2 (red line) and RT3 (green line).
It should be noted that, using the \cite{2007ApJ...657..810D} dust model (which was originally used in RT1 and
RT2) 
instead of that by \cite{2013A&A...558A..62J}
the temperature is systematically higher but the difference between the two 
estimates is never higher 
than $\sim 1$\,K and that temperature profiles follow the same trends.

The dust temperature profile given by our fit follows the same trend as
that estimated using radiative transfer models only for $z \gtrsim 1.5$\,kpc where the
dust temperature is dominated by the thick dust component.
On the contrary, the dust temperature of the thin dust component
drops off more quickly than expected, reaching $T \sim 16$\,K at $\sim 0.5$\,kpc from the
galactic plane.
In the case region X is not affected by the emission of unresolved sources,
this discrepancy would require the presence of two distinct heating sources for the two dust component 
(see Sect.~\ref{sect:heating}).

\subsection{Optical depth estimation}

The other main parameter obtained from the fitting procedure is the optical depth 
at 100 $\mu$m at projected radial distance $h_{\rm X}$, $\tau_{\rm 100}^{\rm e} (h = h_{\rm X})$.
In order to compare this parameter with the face-on central optical depth in the B-band, $\tau_{\rm B}^{\rm f, c}$, 
from the three radiative transfer models cited above (i.e. RT1, RT2 and RT3)
we first convert $\tau_{100}$ in $\tau_{\rm B}$ using the $\tau_{100}/\tau_{\rm B} \simeq 7.5\times 10^{-4}$
from \cite{2013A&A...558A..62J} dust model.
A similar ratio, $\tau_{100}/\tau_{\rm B} \simeq 1.0\times 10^{-3}$, is assumed by 
the \cite{2007ApJ...657..810D} dust model.
% using the ratio
%$\tau_{100}/\tau_{\rm B} \simeq 7.5\times 10^{-4}$ from the \cite{2013A&A...558A..62J} dust model.
%A similar ratio, $\tau_{100}/\tau_{\rm B} \simeq 1.0\times 10^{-3}$, is assumed by 
%the 
The profile of $\tau_{\rm B}^{e} (h = h_{\rm X})$ is shown in the right panel of Fig.~\ref{fig:temp_profile_Pop} 
in black lines
and contributions from the thin and thick dust components are indicated with dotted and dashed
lines, respectively. A zoom of the thick component is illustrated in the inset.

Then, following \cite{1987ApJ...317..637K}, the conversion between the $\tau^{\rm e} (h = h_{\rm X})$ and $\tau^{\rm f,c}$
is given by:
\begin{equation}
\label{eq:conversion}
\tau_{\rm B}^{\rm e} (h = h_{\rm X}) = 18.97\, \tau_{\rm B}^{\rm f, c}.
\end{equation}
We see from Fig.~\ref{fig:temp_profile_Pop} that $\tau^{\rm e}_{\rm B} (h = h_{\rm X}) \simeq 40$ at the
midplane.
Using the conversion factor from Eq.~\ref{eq:conversion} we then obtain $\tau_{\rm B}^{f,c} = 2.1$.
The typical values of $\tau_{\rm B}^{f,c}$ for radiative transfer models of this galaxy are about
a factor two higher: $\tau_{\rm B}^{f,c} = 4$ for \cite{2014ApJ...785L..18S} and \cite{2011A&A...531L..11B} and 
$\tau_{\rm B}^{f,c} = 3.5$ for \cite{2011A&A...527A.109P}.

We have also performed a fit of the optical depth profile to an exponential function of the form:
\begin{equation}
\tau_{\rm B}^{e} (h = h_{\rm X},z) = \tau_{\rm B}^{e} (h = h_{\rm X},z=0) \exp(-z/z_{\rm d,2})
\end{equation}
This fit gives a scaleheight of $z_{d,2} =  (1.44 \pm 0.12)$\,kpc, corresponding to the thick
dust component, in good agreement both with the estimate by \cite{2007ApJ...668..918B}, 
$z_{d,2} =  (1.3 \pm 0.3)$\,kpc
and with that by \cite{2014ApJ...785L..18S}, $z_{d2} \sim [1.2-2.0]$\,kpc.
Also, we calculated that $\sim 2 - 3.3$\% of the mass of dust is present further than 2 kpc from the midplane,
which is consistent with the estimates by \cite{2014ApJ...785L..18S}.

Finally, we compare the value of $A_{\rm V}$ as a function of the vertical distance with the average value
observed by \cite{2010MNRAS.405.1025M} from halo extinction measurements towards distant quasars.
From our observations we find $A_{\rm V} = 0.28, 0.13$ and 0.064 at $z = 1,2$ and 3 kpc, respectively,
in good agreement with values extrapolated from results by Menard et al. (2010): 
$A_{\rm V} = 0.20, 0.11$ and 0.076 at these three distances.

\section{Discussion}
\label{sect:discussion}

As seen in Section \ref{sect:fitting}, the data for NGC 891 require the presence of two dust
components; this result does not depend on the choice of the PSF used in the convolution process.
Furthermore, in order to limit the contamination from isolated point sources or 
from an unresolved disc,
%of point-sources present in the midplane,
we chose to focus on region X for the estimation of the vertical profile.
Fig.~\ref{fig:widths_nol} shows that the scaleheight of the thick dust disc in region X 
is almost constant and undergoes wavelength variations of $\lesssim 30\%$,
while the scaleheight of the thin disc is always smaller than estimates by \cite{1999A&A...344..868X}
and presents a gradient in wavelength and thickens by a factor of $\sim 4$ from
70 $\mu$m to 250 $\mu$m.
A similar trend is seen for all the considered PSFs (see Appendix~\ref{app:plots_scaleheights}) 
and it is therefore independent on this choice.
%The scaleheights of both the thin and the thick discs 
%also undergo variations with wavelength, but they are substantially larger than for
%region X; the thin disc thickens by roughly a factor of 4 from 70 to 250 $\mu$m,
%while the thick disc by a factor of 2. 
%Do not do the plot with only one component, we already show that the X2 is very bad,
%Also, do not show a plot imposing the same scaleheight as Manolis, we just say that the fit is very bad!}

\subsection{Thin dust component}
\label{sect:thincomp}

The observed gradient in the scaleheight of the thin component has strong 
consequences on the dust temperature estimation.
The dust temperature of the thin component drops off very quickly within the
first 0.5 kpc from the midplane making the optical depth estimation more
difficult (see for example the fit to the SED at $z = 0.6$\,kpc in Fig.~\ref{fig:spectra_multi}).
Furthermore, the retrieved dust temperature for $z \sim 0.1 - 0.8$\,kpc is significantly lower than 
the dust temperature estimated from radiative transfer models.

Although our estimates of scaleheights are based on observations of region X,
which is the region least affected by the presence of point sources or of an unresolved disc,
even this region might suffer contamination from the emission of unresolved sources.
This would result in a narrower vertical profile and this effect is likely to be wavelength-dependent,
possibly leading to a gradient in scaleheights like that observed.
%Unfortunately, we are not able to test this hypothesis.

To test this hypothesis, we add to the standard two-component intrinsic 
vertical profile an unresolved super-thin disc with a scaleheight fixed to be equal to half
of the observed image pixel size ($13 - 45$\,pc).
We then convolve the intrinsic vertical profile to the instrument LSF and fit the
resulting profile to the observed vertical profile in region X.

Our fits show that the relative contribution of the super-thin disc with respect 
to the thin and thick components is wavelength-dependent and 
becomes less important going from 70 $\mu$m to 250 $\mu$m;
the relative contribution of the super-thin disc to the observed midplane emission ranges from $\sim 80\%$  
at 70 $\mu$m to $\sim 15\%$ at 250 $\mu$m.
This indicates that the spectrum of the unresolved emission from the 
midplane is not grey but it peaks at wavelengths $\lesssim 70\,\mu$m,
corresponding to a dust temperature $\gtrsim 40$\,K, consistent with 
the expected temperature range of dust heated by nearby hot stars.
The super-thin disc could thus represent the collective emission from dust
heated by nearby hot stars in star-forming regions, like, e.g. the clumpy component 
of the \cite{2000A&A...362..138P,2011A&A...527A.109P} models.
Heating from star-forming regions in the galactic plane of NGC 891
was indeed found to be dominant in the shorter FIR bands \citep{2014A&A...565A...4H}.

\begin{figure}[h!]
\begin{center}
\includegraphics[scale=1]{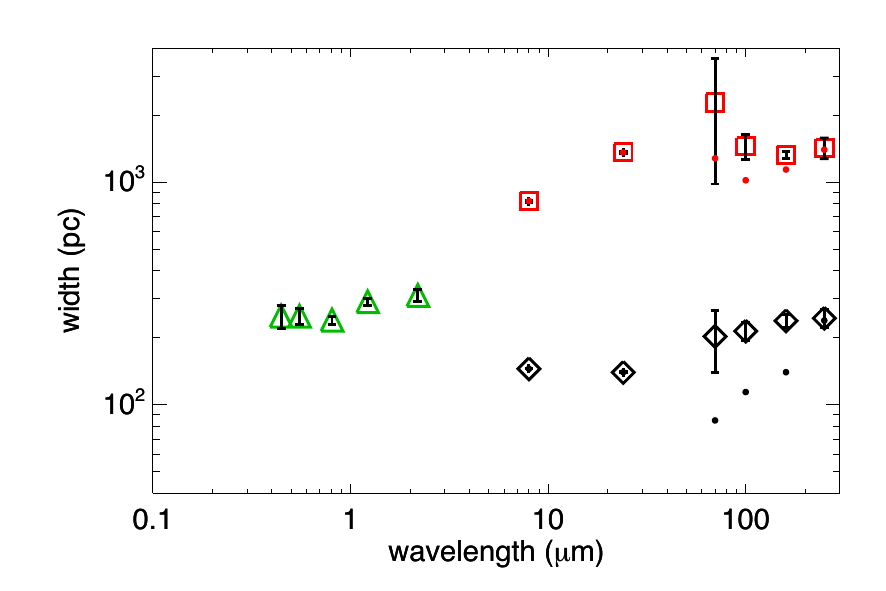}
\caption{Scaleheights of the thin (black diamonds) and thick (red squares) dust components as obtained from 
fitting observations in region X including the super-thin component. 
Green triangles represent the dust scaleheights as estimated by \cite{1999A&A...344..868X}.
Black and red dots are the scaleheights as retrieved in region X using our standard two-component model.}
\label{fig:widths_ps}
\end{center}
\end{figure}

In Fig.~\ref{fig:widths_ps} we show the scaleheights of the thin (black diamonds) 
and thick (red squares) components
obtained fitting observations in region X with this three-component model
and compared them with the scaleheights obtained using the standard two-component
model (black and red dots).
The addition of the super-thin component leads to a milder gradient
in the scaleheight of the thin component for wavelengths in the range $70 - 250\,\mu$m.
This has consequences on the dust temperature estimation leading to a flatter
temperature profile as a function of the vertical height above the midplane, more
similarly to that expected from RT models.
%\ref{sect:dust_temp_est}

On the other hand, we find that the 8 and 24 $\mu$m data are best fit with a
null contribution from the super-thin disc, thus not leading to any changes
in the scaleheight of the thin and thick components.
If we admit that $70 - 250\,\mu$m data are affected by unresolved emission even in
region X while 8 and 24 $\mu$m data are not (or to a low extent), then the estimate
of the fraction of small grains destroyed at high galactic latitudes ($\sim 2.5$\,kpc above the galactic disc)
given in Section~\ref{sect:dust_modelling} will represent a lower limit.

However, the inclusion of a super-thin disc to the standard two-component vertical profile
adds complexity to the model and leads to overfitting the data.
In fact, the resolution of the available instruments is not sufficient to probe 
scales of tens of parsecs at the distance of this galaxy.
We have therefore presented the results from fitting a three-component model
only qualitatively in order to show the consequences of the presence
of unresolved sources in the midplane.

%1. the relative contribution goes down from 70 to 250 um.
%which indicates that point sources have a colour. (DONE!)

%2. the scale height of the thin component does show a milder gradient
%the flatter scale height dependence is reflected in the dust temperature,
%which tends to follow more the RT models. (DONE!)

%3. while no change has been found for MIR wavelengths.
%More destruction of small grains.

%4. however, these results have to be considered only qualitatively since we are 
%over fitting the data with a too complex model and we are not
%able to prefer this scenario to the one described above.

%This operation is however degenerate and the results must be considered qualitative
%since we are overfitting the data and with the resolution of these instruments 
%we are not able to disentangle between these two models.}
 
%Furthermore, it has to be noted that the ratio $\tau_{100} / \tau_{\rm B}$ is not well constrained observationally.
%For example this quantity can be estimated from observations in our Galaxy in this way:
%\begin{equation}
%\frac{\tau_{100}}{\tau_{\rm B}} = \frac{\tau_{100}}{N_{\rm H}} \frac{N_{\rm H}}{E(B-V)} \frac{E(B-V)}{\tau_{\rm B}},
%\end{equation}
%where $N_{\rm H}$ is the hydrogen column density and $E(B-V) = A_{\rm B} - A_{\rm V}$.
%Recent observations from \cite{2014A&A...571A..11P} resulted in $\frac{\tau_{100}}{N_{\rm H}} = 2.1 \times 10^{-25}$\,cm$^2$
%and \cite{2014ApJ...780...10L} estimated $\frac{N_{\rm H}}{E(B-V)} \simeq 8.3 \times 10^{21}$\,cm$^{-2}$.
%Then, using $R_{\rm V} = 3.1$, we obtain $\frac{\tau_{100}}{\tau_{\rm B}} \sim 1.7 \times 10^{-4}$, which is $\sim 4.4$ and
%$\sim 5.9$ times lower than values assumed by the \cite{2013A&A...558A..62J} and \cite{2007ApJ...657..810D} dust models, respectively.
%However, this does not affect our fitting procedure and in particular our estimates of the fraction of dust present in the halo
%of the galaxy or of the scaleheight of the thick dust component.

\subsection{Heating sources}
\label{sect:heating}

In the case where region X was not highly affected by unresolved emission the big difference in
dust temperature between the thin and thick components requires two
different heating sources.
The thin component is likely heated by the galactic radiation field, which would drop 
off much more quickly than expected by radiative transfer models.
On the other hand, the thick dust component might be heated by the presence
of a hot halo gas or from strong radiation from nearby hot evolved stars.

Assuming that the hot halo gas is distributed as observed by \cite{2013ApJ...762...12H}
and following the method by \cite{2013A&A...556A...6B} we computed the dust emission
in the case of collisional heating only.
We obtained a dust temperature $T_{\rm dust} = 19, 18.2$\,and 17.4 K at $z = 1,2$\,and 3 kpc from the midplane, respectively,
which is not enough to explain the observed high dust temperature.

Another possibility comes from the strong radiation field from evolved
stars present in the halo.
In fact, accordingly to the model by \cite{2011MNRAS.415.2182F}, hot low-mass evolved stars are 
present in the halo of NGC 891 and dominate the ionisation of the halo gas and 
could possibly heat nearby dust to high temperatures.
A similar source of heating was considered by \cite{2007ApJ...668..918B} in order to explain 
the observed emission from halo dust.
This model is also consistent with HST observations from \cite{2009MNRAS.395..126I}, which 
show a thick stellar component with a vertical scaleheight $z_{\rm s} = (1.44 \pm 0.03)$\,kpc.

Furthermore, observations of the H{\sc i} gas distribution from \cite{2007AJ....134.1019O}
show the presence of an H{\sc i} halo extending out to about 14 kpc from the midplane 
in the northeast, southeast, and southwest quadrants while in the northwest quadrant 
a flare extending out to $\sim 22$\,kpc is observed.
A few years later \cite{2011AJ....141...48Y}, using Very Large Array observations
detected a thin and a thick H{\sc i} components.
They modelled the thick distribution with a Gaussian function 
with a FWHM of $\sim 2$\,kpc which roughly translates into an exponential decay with a scaleheight of
$\sim 1.4$\,kpc. 
Moreover, this galaxy has been mapped in the J=2-1 and 1-0 lines
of $^{12}$CO with the IRAM 30-m telescope by \cite{1992A&A...266...21G}
and in the J=1-0 line with the Nobeyama 45-m telescope by \cite{1993PASJ...45..139S}.
These two teams detected a `halo' of molecular gas $\geqslant 7$\,kpc in radius and 
$2-3$\,kpc thick. This component is very weak in the 1-0 line and seems to be 
relatively hot and not too optically thick.
This means that dust, H{\sc i} gas, molecular gas and stars follow a similar distribution in the halo and 
that dust might be embedded in atomic / molecular gas and heated by this secondary 
distribution of stars.
%{\bf Add first HI observations by Oosterloo+07, this comes first, with its warp.
%Then Yim for HI and the scaleheight and then CO by ... 
%And finally, dust would be embedded in HI or molecular gas clouds and heated by...
%Also, a large CO and HI distributions have been observed by ...
%ADD CO and HI observations by \cite{1992A&A...266...21G} and \cite{1993PASJ...45..139S} 
%and .
%And maybe add sth in the formation of the halo, the warp is not seen in dust emission, this might indicate...}

On the contrary, if we admit that, even in region X there might be some contamination
due to the presence of a super-thin disc 
(as suggested by our three-component model analysis in Sect~\ref{sect:thincomp}),
then the gradient in scaleheight of the thin component
could be heavily smoothed.
This would lead to a flatter profile in temperature, possibly similar to
that of the thick component.
This scenario then does not require two distinct sources of heating and
the galactic radiation field as modeled by various radiative transfer models would be
sufficient to explain the high dust temperature in the halo of the galaxy.

\subsection{The origin of extra-planar dust}

We have detected dust emission at $z \geqslant 2$\,kpc from the galactic disc.
However, the evidence for extra-planar dust is not limited to NGC 891.
Dusty clouds have been observed in extinction
up to $z \sim 2$ in the thick discs of the majority
of the edge-on galaxies (\citealt{1997AJ....114.2463H, 1999AJ....117.2077H, 
2000AJ....119..644H,2004AJ....128..674R,
2004AJ....128..662T,2005ASPC..331..287H,2012EAS....56..291H}).
The optical images of these galaxies often show  
filamentary structures perpendicular to the galactic disc, linking the thin and thick discs.
Also, extended (up to $\sim 6$\,kpc) mid-infrared 
aromatic feature emission is seen around a number of galaxies (see e.g. 
\citealt{2006ApJ...642L.127E,2006A&A...445..123I,2013ApJ...774..126M}).
%Also Menard et al. 2010 show the evidence of dust even further in the IGM up to 1Mpc from galaxies.
%Furthermore, direct optical images of edge-on galaxies often show filamentary structures
%perpendicular to the galactic disc, extending a few kpc from it, linking the thin and thick discs
%(\citealt{1997AJ....114.2463H, 1999AJ....117.2077H, 2000AJ....119..644H,2004AJ....128..674R,
%2004AJ....128..662T,2005ASPC..331..287H}).

The presence of dust in galactic haloes can be the result of several important phenomena.
We identified at least three mechanisms, possibly able to explain the presence of dust
at these distances above the disc.
1) Dust can be expelled due to galactic fountains or stellar feedback (local winds).
Observational and theoretical evidence show that small grains are partly destroyed 
during coupling with shocks and galactic fountains 
(\citealt{2006ApJ...642L.141B,2006ApJ...652L..33W,2010A&A...510A..36M,2014A&A...570A..32B}),
in agreement with what we found.
2) Slow global winds (e.g. cosmic ray driven winds) could also explain the presence of dust in the galactic halo. 
These winds are slowly and continuously emitted by spiral galaxies pushing dust grains to high 
galactic latitudes.
Also, the expansion of these winds lead to a rapid drop in the gas temperature, therefore 
reducing grain sputtering (\citealt{2000A&A...354..480P}).
3) Another possibility comes from the radiation pressure. In fact, \cite{1991ApJ...381..137F}
show that this mechanism might displace dust up to several kpc from the galactic disc.
4) Finally, a significant contribution to the observed dust might come from dust formed in-situ
around AGB stars present in the halo (\citealt{2012EAS....56..291H}).
{\rm However, this scenario is not very likely since it would not explain the grain erosion 
above the galactic midplane.}

A significant amount of dust is observed even further out in the IGM
around galaxies (\citealt{1994AJ....108.1619Z,2010MNRAS.405.1025M}).
The thick dusty disc described here would be an interface from which dust is ejected 
into the IGM.

\section{Summary}
\label{sect:summary}

We have analysed deep PACS data, together with IRAC, MIPS and SPIRE data and we 
have detected dust emission from the halo of NGC 891.
Our main findings are summarised here.
\begin{itemize}
\item Dust emission at all wavelengths is best fit by a double exponential profile revealing the 
presence of two dust components: a thin and a thick disc.
\item The measured scaleheight of the thin disc shows a gradient through the FIR.
This can be possibly explained by a dramatic drop in dust temperature within the
thin disc scaleheight.
Alternatively, contamination from a super-thin disc may also 
produce this effect and reconcile the observations with standard RT models for this galaxy.
%point sources present in the galactic disc 
\item We obtain a precise measurement of the thick halo scaleheight, $z_{\rm d,2} = (1.44\pm 0.12)$\,kpc,
which is in agreement with previous estimates (\citealt{2007ApJ...668..918B,2007A&A...471L...1K,
2009MNRAS.395...97W,2014ApJ...785L..18S}).
\item We find that a non-negligible fraction ($2 - 3.3$\%) of the mass of dust is located further than 2 kpc from
the midplane, in accord with findings by \cite{2014ApJ...785L..18S}.
Furthermore, the relative abundance of small grains with respect to large grains
drops from $m_{\rm SG} / m_{\rm LG} = 
(6.2\pm 0.1)\times 10^{-2}$ in the galactic disc to 
$m_{\rm SG} / m_{\rm LG} = (3.9\pm 0.1)\times 10^{-2}$ at a height of $\sim 2.5$\,kpc
above the plane suggesting that dust is hit by interstellar shocks or galactic fountains 
and entrained together with gas to a height of a few kpc from the midplane.
\item The detected halo dust component is found to have a similar extent of the
atomic / molecular gas distribution and of the thick stellar disc described by \cite{2009MNRAS.395..126I}.
The thick stellar disc can therefore be a possible source of heating. 
%and also a possible source of
%dust.
\end{itemize}

\begin{acknowledgements}
M. B. wishes to acknowledge A. Abergel (IAS, France) and H. Roussel (IAP, France) for interesting discussions on the modelling 
of {\it Herschel} PSFs and in particular H. Roussel for having performed data reduction for the images of Vesta and Mars.
We would like to thank the referee, C. C. Popescu, for her very helpful comments.
The research leading to these results has received funding from the 
European Research Council under the European Union (FP/2007-2013) / ERC
Grant Agreement n. 306476.
\end{acknowledgements}

\bibliographystyle{aa}
\bibliography{bib_n891}{}

\begin{thebibliography}{67}
\expandafter\ifx\csname natexlab\endcsname\relax\def\natexlab#1{#1}\fi

\bibitem[{{Alton} {et~al.}(2004){Alton}, {Xilouris}, {Misiriotis}, {Dasyra}, \&
  {Dumke}}]{2004A&A...425..109A}
{Alton}, P.~B., {Xilouris}, E.~M., {Misiriotis}, A., {Dasyra}, K.~M., \&
  {Dumke}, M. 2004, \aap, 425, 109

\bibitem[{{Aniano} {et~al.}(2011){Aniano}, {Draine}, {Gordon}, \&
  {Sandstrom}}]{2011PASP..123.1218A}
{Aniano}, G., {Draine}, B.~T., {Gordon}, K.~D., \& {Sandstrom}, K. 2011, \pasp,
  123, 1218

\bibitem[{{Baes} {et~al.}(2010){Baes}, {Fritz}, {Gadotti}, {Smith}, {Dunne},
  {da Cunha}, {Amblard}, {Auld}, {Bendo}, {Bonfield}, {Burgarella},
  {Buttiglione}, {Cava}, {Clements}, {Cooray}, {Dariush}, {de Zotti}, {Dye},
  {Eales}, {Frayer}, {Gonzalez-Nuevo}, {Herranz}, {Ibar}, {Ivison}, {Lagache},
  {Leeuw}, {Lopez-Caniego}, {Jarvis}, {Maddox}, {Negrello}, {Micha{\l}owski},
  {Pascale}, {Pohlen}, {Rigby}, {Rodighiero}, {Samui}, {Serjeant}, {Temi},
  {Thompson}, {van der Werf}, {Verma}, \& {Vlahakis}}]{2010A&A...518L..39B}
{Baes}, M., {Fritz}, J., {Gadotti}, D.~A., {et~al.} 2010, \aap, 518, L39

\bibitem[{{Balog} {et~al.}(2014){Balog}, {Muzerolle}, {Flaherty}, {Detre},
  {Bouwmann}, {Furlan}, {Gutermuth}, {Juhasz}, {Bally}, {Nielbock}, {Klaas},
  {Krause}, {Henning}, \& {Marton}}]{2014ApJ...789L..38B}
{Balog}, Z., {Muzerolle}, J., {Flaherty}, K., {et~al.} 2014, \apjl, 789, L38

\bibitem[{{Bendo} {et~al.}(2012){Bendo}, {Galliano}, \&
  {Madden}}]{2012MNRAS.423..197B}
{Bendo}, G.~J., {Galliano}, F., \& {Madden}, S.~C. 2012, \mnras, 423, 197

\bibitem[{{Bianchi}(2007)}]{2007A&A...471..765B}
{Bianchi}, S. 2007, \aap, 471, 765

\bibitem[{{Bianchi}(2008)}]{2008A&A...490..461B}
{Bianchi}, S. 2008, \aap, 490, 461

\bibitem[{{Bianchi} \& {Xilouris}(2011)}]{2011A&A...531L..11B}
{Bianchi}, S. \& {Xilouris}, E.~M. 2011, \aap, 531, L11

\bibitem[{{Bocchio}(2014)}]{mythesis}
{Bocchio}, M. 2014, PhD thesis, Universit{\'e} Paris Sud

\bibitem[{{Bocchio} {et~al.}(2014){Bocchio}, {Jones}, \&
  {Slavin}}]{2014A&A...570A..32B}
{Bocchio}, M., {Jones}, A.~P., \& {Slavin}, J.~D. 2014, \aap, 570, A32

\bibitem[{{Bocchio} {et~al.}(2013){Bocchio}, {Jones}, {Verstraete}, {Xilouris},
  {Micelotta}, \& {Bianchi}}]{2013A&A...556A...6B}
{Bocchio}, M., {Jones}, A.~P., {Verstraete}, L., {et~al.} 2013, \aap, 556, A6

\bibitem[{{Bocchio} {et~al.}(2012){Bocchio}, {Micelotta}, {Gautier}, \&
  {Jones}}]{2012A&A...545A.124B}
{Bocchio}, M., {Micelotta}, E.~R., {Gautier}, A.-L., \& {Jones}, A.~P. 2012,
  \aap, 545, A124

\bibitem[{{Borkowski} {et~al.}(2006){Borkowski}, {Williams}, {Reynolds},
  {Blair}, {Ghavamian}, {Sankrit}, {Hendrick}, {Long}, {Raymond}, {Smith},
  {Points}, \& {Winkler}}]{2006ApJ...642L.141B}
{Borkowski}, K.~J., {Williams}, B.~J., {Reynolds}, S.~P., {et~al.} 2006, \apjl,
  642, L141

\bibitem[{{Burgdorf} {et~al.}(2007){Burgdorf}, {Ashby}, \&
  {Williams}}]{2007ApJ...668..918B}
{Burgdorf}, M., {Ashby}, M.~L.~N., \& {Williams}, R. 2007, \apj, 668, 918

\bibitem[{{Ciesla} {et~al.}(2012){Ciesla}, {Boselli}, {Smith}, {Bendo},
  {Cortese}, {Eales}, {Bianchi}, {Boquien}, {Buat}, {Davies}, {Pohlen},
  {Zibetti}, {Baes}, {Cooray}, {De Looze}, {di Serego Alighieri}, {Galametz},
  {Gomez}, {Lebouteiller}, {Madden}, {Pappalardo}, {Remy}, {Spinoglio},
  {Vaccari}, {Auld}, \& {Clements}}]{2012A&A...543A.161C}
{Ciesla}, L., {Boselli}, A., {Smith}, M.~W.~L., {et~al.} 2012, \aap, 543, A161

\bibitem[{{Cortese} {et~al.}(2014){Cortese}, {Fritz}, {Bianchi}, {Boselli},
  {Ciesla}, {Bendo}, {Boquien}, {Roussel}, {Baes}, {Buat}, {Clemens}, {Cooray},
  {Cormier}, {Davies}, {De Looze}, {Eales}, {Fuller}, {Hunt}, {Madden},
  {Munoz-Mateos}, {Pappalardo}, {Pierini}, {R{\'e}my-Ruyer}, {Sauvage}, {di
  Serego Alighieri}, {Smith}, {Spinoglio}, {Vaccari}, \&
  {Vlahakis}}]{2014MNRAS.440..942C}
{Cortese}, L., {Fritz}, J., {Bianchi}, S., {et~al.} 2014, \mnras, 440, 942

\bibitem[{{De Geyter} {et~al.}(2014){De Geyter}, {Baes}, {Camps}, {Fritz}, {De
  Looze}, {Hughes}, {Viaene}, \& {Gentile}}]{2014MNRAS.441..869D}
{De Geyter}, G., {Baes}, M., {Camps}, P., {et~al.} 2014, \mnras, 441, 869

\bibitem[{{De Looze} {et~al.}(2012){De Looze}, {Baes}, {Bendo}, {Ciesla},
  {Cortese}, {de Geyter}, {Groves}, {Boquien}, {Boselli}, {Brondeel}, {Cooray},
  {Eales}, {Fritz}, {Galliano}, {Gentile}, {Gordon}, {Hony}, {Law}, {Madden},
  {Sauvage}, {Smith}, {Spinoglio}, \& {Verstappen}}]{2012MNRAS.427.2797D}
{De Looze}, I., {Baes}, M., {Bendo}, G.~J., {et~al.} 2012, \mnras, 427, 2797

\bibitem[{{Draine} \& {Li}(2007)}]{2007ApJ...657..810D}
{Draine}, B.~T. \& {Li}, A. 2007, \apj, 657, 810

\bibitem[{{Ebert} {et~al.}(2002){Ebert}, {Musgrave}, {Peachey}, {Perlin}, \&
  {Worley}}]{smooth}
{Ebert}, D.~S., {Musgrave}, F.~K., {Peachey}, D., {Perlin}, K., \& {Worley}, S.
  2002, {Texturing and Modeling} (Morgan Kaufmann; 3 edition)

\bibitem[{{Engelbracht} {et~al.}(2006){Engelbracht}, {Kundurthy}, {Gordon},
  {Rieke}, {Kennicutt}, {Smith}, {Regan}, {Makovoz}, {Sosey}, {Draine},
  {Helou}, {Armus}, {Calzetti}, {Meyer}, {Bendo}, {Walter}, {Hollenbach},
  {Cannon}, {Murphy}, {Dale}, {Buckalew}, \& {Sheth}}]{2006ApJ...642L.127E}
{Engelbracht}, C.~W., {Kundurthy}, P., {Gordon}, K.~D., {et~al.} 2006, \apjl,
  642, L127

\bibitem[{{Ferrara} {et~al.}(1991){Ferrara}, {Ferrini}, {Barsella}, \&
  {Franco}}]{1991ApJ...381..137F}
{Ferrara}, A., {Ferrini}, F., {Barsella}, B., \& {Franco}, J. 1991, \apj, 381,
  137

\bibitem[{{Flores-Fajardo} {et~al.}(2011){Flores-Fajardo}, {Morisset},
  {Stasi{\'n}ska}, \& {Binette}}]{2011MNRAS.415.2182F}
{Flores-Fajardo}, N., {Morisset}, C., {Stasi{\'n}ska}, G., \& {Binette}, L.
  2011, \mnras, 415, 2182

\bibitem[{{Garcia-Burillo} {et~al.}(1992){Garcia-Burillo}, {Guelin},
  {Cernicharo}, \& {Dahlem}}]{1992A&A...266...21G}
{Garcia-Burillo}, S., {Guelin}, M., {Cernicharo}, J., \& {Dahlem}, M. 1992,
  \aap, 266, 21

\bibitem[{{Griffin} {et~al.}(2010){Griffin}, {Abergel}, {Abreu}, {Ade},
  {Andr{\'e}}, {Augueres}, {Babbedge}, {Bae}, {Baillie}, {Baluteau}, {Barlow},
  {Bendo}, {Benielli}, {Bock}, {Bonhomme}, {Brisbin}, {Brockley-Blatt},
  {Caldwell}, {Cara}, {Castro-Rodriguez}, {Cerulli}, {Chanial}, {Chen},
  {Clark}, {Clements}, {Clerc}, {Coker}, {Communal}, {Conversi}, {Cox},
  {Crumb}, {Cunningham}, {Daly}, {Davis}, {de Antoni}, {Delderfield}, {Devin},
  {di Giorgio}, {Didschuns}, {Dohlen}, {Donati}, {Dowell}, {Dowell}, {Duband},
  {Dumaye}, {Emery}, {Ferlet}, {Ferrand}, {Fontignie}, {Fox}, {Franceschini},
  {Frerking}, {Fulton}, {Garcia}, {Gastaud}, {Gear}, {Glenn}, {Goizel},
  {Griffin}, {Grundy}, {Guest}, {Guillemet}, {Hargrave}, {Harwit}, {Hastings},
  {Hatziminaoglou}, {Herman}, {Hinde}, {Hristov}, {Huang}, {Imhof}, {Isaak},
  {Israelsson}, {Ivison}, {Jennings}, {Kiernan}, {King}, {Lange}, {Latter},
  {Laurent}, {Laurent}, {Leeks}, {Lellouch}, {Levenson}, {Li}, {Li},
  {Lilienthal}, {Lim}, {Liu}, {Lu}, {Madden}, {Mainetti}, {Marliani}, {McKay},
  {Mercier}, {Molinari}, {Morris}, {Moseley}, {Mulder}, {Mur}, {Naylor},
  {Nguyen}, {O'Halloran}, {Oliver}, {Olofsson}, {Olofsson}, {Orfei}, {Page},
  {Pain}, {Panuzzo}, {Papageorgiou}, {Parks}, {Parr-Burman}, {Pearce},
  {Pearson}, {P{\'e}rez-Fournon}, {Pinsard}, {Pisano}, {Podosek}, {Pohlen},
  {Polehampton}, {Pouliquen}, {Rigopoulou}, {Rizzo}, {Roseboom}, {Roussel},
  {Rowan-Robinson}, {Rownd}, {Saraceno}, {Sauvage}, {Savage}, {Savini},
  {Sawyer}, {Scharmberg}, {Schmitt}, {Schneider}, {Schulz}, {Schwartz},
  {Shafer}, {Shupe}, {Sibthorpe}, {Sidher}, {Smith}, {Smith}, {Smith},
  {Spencer}, {Stobie}, {Sudiwala}, {Sukhatme}, {Surace}, {Stevens}, {Swinyard},
  {Trichas}, {Tourette}, {Triou}, {Tseng}, {Tucker}, {Turner}, {Vaccari},
  {Valtchanov}, {Vigroux}, {Virique}, {Voellmer}, {Walker}, {Ward}, {Waskett},
  {Weilert}, {Wesson}, {White}, {Whitehouse}, {Wilson}, {Winter}, {Woodcraft},
  {Wright}, {Xu}, {Zavagno}, {Zemcov}, {Zhang}, \&
  {Zonca}}]{2010A&A...518L...3G}
{Griffin}, M.~J., {Abergel}, A., {Abreu}, A., {et~al.} 2010, \aap, 518, L3

\bibitem[{{Helou} {et~al.}(2004){Helou}, {Roussel}, {Appleton}, {Frayer},
  {Stolovy}, {Storrie-Lombardi}, {Hurt}, {Lowrance}, {Makovoz}, {Masci},
  {Surace}, {Gordon}, {Alonso-Herrero}, {Engelbracht}, {Misselt}, {Rieke},
  {Rieke}, {Willner}, {Pahre}, {Ashby}, {Fazio}, \&
  {Smith}}]{2004ApJS..154..253H}
{Helou}, G., {Roussel}, H., {Appleton}, P., {et~al.} 2004, \apjs, 154, 253

\bibitem[{{Hodges-Kluck} \& {Bregman}(2014)}]{2014ApJ...789..131H}
{Hodges-Kluck}, E. \& {Bregman}, J.~N. 2014, \apj, 789, 131

\bibitem[{{Hodges-Kluck} \& {Bregman}(2013)}]{2013ApJ...762...12H}
{Hodges-Kluck}, E.~J. \& {Bregman}, J.~N. 2013, \apj, 762, 12

\bibitem[{{Howk}(2005)}]{2005ASPC..331..287H}
{Howk}, J.~C. 2005, in Astronomical Society of the Pacific Conference Series,
  Vol. 331, Extra-Planar Gas, ed. R.~{Braun}, 287

\bibitem[{{Howk}(2012)}]{2012EAS....56..291H}
{Howk}, J.~C. 2012, in EAS Publications Series, Vol.~56, EAS Publications
  Series, ed. M.~A. {de Avillez}, 291--298

\bibitem[{{Howk} \& {Savage}(1997)}]{1997AJ....114.2463H}
{Howk}, J.~C. \& {Savage}, B.~D. 1997, \aj, 114, 2463

\bibitem[{{Howk} \& {Savage}(1999)}]{1999AJ....117.2077H}
{Howk}, J.~C. \& {Savage}, B.~D. 1999, \aj, 117, 2077

\bibitem[{{Howk} \& {Savage}(2000)}]{2000AJ....119..644H}
{Howk}, J.~C. \& {Savage}, B.~D. 2000, \aj, 119, 644

\bibitem[{{Hughes} {et~al.}(2014){Hughes}, {Baes}, {Fritz}, {Smith}, {Parkin},
  {Gentile}, {Bendo}, {Wilson}, {Allaert}, {Bianchi}, {De Looze}, {Verstappen},
  {Viaene}, {Boquien}, {Boselli}, {Clements}, {Davies}, {Galametz}, {Madden},
  {R{\'e}my-Ruyer}, \& {Spinoglio}}]{2014A&A...565A...4H}
{Hughes}, T.~M., {Baes}, M., {Fritz}, J., {et~al.} 2014, \aap, 565, A4

\bibitem[{{Ibata} {et~al.}(2009){Ibata}, {Mouhcine}, \&
  {Rejkuba}}]{2009MNRAS.395..126I}
{Ibata}, R., {Mouhcine}, M., \& {Rejkuba}, M. 2009, \mnras, 395, 126

\bibitem[{{Irwin} \& {Madden}(2006)}]{2006A&A...445..123I}
{Irwin}, J.~A. \& {Madden}, S.~C. 2006, \aap, 445, 123

\bibitem[{{Jones} {et~al.}(2013){Jones}, {Fanciullo}, {K{\"o}hler},
  {Verstraete}, {Guillet}, {Bocchio}, \& {Ysard}}]{2013A&A...558A..62J}
{Jones}, A.~P., {Fanciullo}, L., {K{\"o}hler}, M., {et~al.} 2013, \aap, 558,
  A62

\bibitem[{{Kamphuis} {et~al.}(2007){Kamphuis}, {Holwerda}, {Allen}, {Peletier},
  \& {van der Kruit}}]{2007A&A...471L...1K}
{Kamphuis}, P., {Holwerda}, B.~W., {Allen}, R.~J., {Peletier}, R.~F., \& {van
  der Kruit}, P.~C. 2007, \aap, 471, L1

\bibitem[{{Kylafis} \& {Bahcall}(1987)}]{1987ApJ...317..637K}
{Kylafis}, N.~D. \& {Bahcall}, J.~N. 1987, \apj, 317, 637

\bibitem[{{Makovoz} \& {Marleau}(2005)}]{2005PASP..117.1113M}
{Makovoz}, D. \& {Marleau}, F.~R. 2005, \pasp, 117, 1113

\bibitem[{{Mathis} {et~al.}(1983){Mathis}, {Mezger}, \&
  {Panagia}}]{1983A&A...128..212M}
{Mathis}, J.~S., {Mezger}, P.~G., \& {Panagia}, N. 1983, \aap, 128, 212

\bibitem[{{McCormick} {et~al.}(2013){McCormick}, {Veilleux}, \&
  {Rupke}}]{2013ApJ...774..126M}
{McCormick}, A., {Veilleux}, S., \& {Rupke}, D.~S.~N. 2013, \apj, 774, 126

\bibitem[{{M{\'e}nard} {et~al.}(2010){M{\'e}nard}, {Scranton}, {Fukugita}, \&
  {Richards}}]{2010MNRAS.405.1025M}
{M{\'e}nard}, B., {Scranton}, R., {Fukugita}, M., \& {Richards}, G. 2010,
  \mnras, 405, 1025

\bibitem[{{Micelotta} {et~al.}(2010){Micelotta}, {Jones}, \&
  {Tielens}}]{2010A&A...510A..36M}
{Micelotta}, E.~R., {Jones}, A.~P., \& {Tielens}, A.~G.~G.~M. 2010, \aap, 510,
  A36

\bibitem[{{Oosterloo} {et~al.}(2007){Oosterloo}, {Fraternali}, \&
  {Sancisi}}]{2007AJ....134.1019O}
{Oosterloo}, T., {Fraternali}, F., \& {Sancisi}, R. 2007, \aj, 134, 1019

\bibitem[{{Ott}(2010)}]{2010ASPC..434..139O}
{Ott}, S. 2010, in Astronomical Society of the Pacific Conference Series, Vol.
  434, Astronomical Data Analysis Software and Systems XIX, ed. Y.~{Mizumoto},
  K.-I. {Morita}, \& M.~{Ohishi}, 139

\bibitem[{{Pilbratt} {et~al.}(2010){Pilbratt}, {Riedinger}, {Passvogel},
  {Crone}, {Doyle}, {Gageur}, {Heras}, {Jewell}, {Metcalfe}, {Ott}, \&
  {Schmidt}}]{2010A&A...518L...1P}
{Pilbratt}, G.~L., {Riedinger}, J.~R., {Passvogel}, T., {et~al.} 2010, \aap,
  518, L1

\bibitem[{{Poglitsch} {et~al.}(2010){Poglitsch}, {Waelkens}, {Geis},
  {Feuchtgruber}, {Vandenbussche}, {Rodriguez}, {Krause}, {Renotte}, {van
  Hoof}, {Saraceno}, {Cepa}, {Kerschbaum}, {Agn{\`e}se}, {Ali}, {Altieri},
  {Andreani}, {Augueres}, {Balog}, {Barl}, {Bauer}, {Belbachir}, {Benedettini},
  {Billot}, {Boulade}, {Bischof}, {Blommaert}, {Callut}, {Cara}, {Cerulli},
  {Cesarsky}, {Contursi}, {Creten}, {De Meester}, {Doublier}, {Doumayrou},
  {Duband}, {Exter}, {Genzel}, {Gillis}, {Gr{\"o}zinger}, {Henning},
  {Herreros}, {Huygen}, {Inguscio}, {Jakob}, {Jamar}, {Jean}, {de Jong},
  {Katterloher}, {Kiss}, {Klaas}, {Lemke}, {Lutz}, {Madden}, {Marquet},
  {Martignac}, {Mazy}, {Merken}, {Montfort}, {Morbidelli}, {M{\"u}ller},
  {Nielbock}, {Okumura}, {Orfei}, {Ottensamer}, {Pezzuto}, {Popesso},
  {Putzeys}, {Regibo}, {Reveret}, {Royer}, {Sauvage}, {Schreiber}, {Stegmaier},
  {Schmitt}, {Schubert}, {Sturm}, {Thiel}, {Tofani}, {Vavrek}, {Wetzstein},
  {Wieprecht}, \& {Wiezorrek}}]{2010A&A...518L...2P}
{Poglitsch}, A., {Waelkens}, C., {Geis}, N., {et~al.} 2010, \aap, 518, L2

\bibitem[{{Popescu} {et~al.}(2000{\natexlab{a}}){Popescu}, {Misiriotis},
  {Kylafis}, {Tuffs}, \& {Fischera}}]{2000A&A...362..138P}
{Popescu}, C.~C., {Misiriotis}, A., {Kylafis}, N.~D., {Tuffs}, R.~J., \&
  {Fischera}, J. 2000{\natexlab{a}}, \aap, 362, 138

\bibitem[{{Popescu} \& {Tuffs}(2013)}]{2013MNRAS.436.1302P}
{Popescu}, C.~C. \& {Tuffs}, R.~J. 2013, \mnras, 436, 1302

\bibitem[{{Popescu} {et~al.}(2011){Popescu}, {Tuffs}, {Dopita}, {Fischera},
  {Kylafis}, \& {Madore}}]{2011A&A...527A.109P}
{Popescu}, C.~C., {Tuffs}, R.~J., {Dopita}, M.~A., {et~al.} 2011, \aap, 527,
  A109

\bibitem[{{Popescu} {et~al.}(2000{\natexlab{b}}){Popescu}, {Tuffs}, {Fischera},
  \& {V{\"o}lk}}]{2000A&A...354..480P}
{Popescu}, C.~C., {Tuffs}, R.~J., {Fischera}, J., \& {V{\"o}lk}, H.
  2000{\natexlab{b}}, \aap, 354, 480

\bibitem[{{Rossa} {et~al.}(2004){Rossa}, {Dettmar}, {Walterbos}, \&
  {Norman}}]{2004AJ....128..674R}
{Rossa}, J., {Dettmar}, R.-J., {Walterbos}, R.~A.~M., \& {Norman}, C.~A. 2004,
  \aj, 128, 674

\bibitem[{{Roussel}(2013)}]{2013PASP..125.1126R}
{Roussel}, H. 2013, \pasp, 125, 1126

\bibitem[{{Seon} \& {Witt}(2012)}]{2012IAUS..284..135S}
{Seon}, K.-I. \& {Witt}, A.~N. 2012, in IAU Symposium, Vol. 284, IAU Symposium,
  ed. R.~J. {Tuffs} \& C.~C. {Popescu}, 135--137

\bibitem[{{Seon} {et~al.}(2014){Seon}, {Witt}, {Shinn}, \&
  {Kim}}]{2014ApJ...785L..18S}
{Seon}, K.-i., {Witt}, A.~N., {Shinn}, J.-h., \& {Kim}, I.-j. 2014, \apjl, 785,
  L18

\bibitem[{{Sofue} \& {Nakai}(1993)}]{1993PASJ...45..139S}
{Sofue}, Y. \& {Nakai}, N. 1993, \pasj, 45, 139

\bibitem[{{Strickland} {et~al.}(2004){Strickland}, {Heckman}, {Colbert},
  {Hoopes}, \& {Weaver}}]{2004ApJS..151..193S}
{Strickland}, D.~K., {Heckman}, T.~M., {Colbert}, E.~J.~M., {Hoopes}, C.~G., \&
  {Weaver}, K.~A. 2004, \apjs, 151, 193

\bibitem[{{Thompson} {et~al.}(2004){Thompson}, {Howk}, \&
  {Savage}}]{2004AJ....128..662T}
{Thompson}, T.~W.~J., {Howk}, J.~C., \& {Savage}, B.~D. 2004, \aj, 128, 662

\bibitem[{{Verstappen} {et~al.}(2013){Verstappen}, {Fritz}, {Baes}, {Smith},
  {Allaert}, {Bianchi}, {Blommaert}, {De Geyter}, {De Looze}, {Gentile},
  {Gordon}, {Holwerda}, {Viaene}, \& {Xilouris}}]{2013A&A...556A..54V}
{Verstappen}, J., {Fritz}, J., {Baes}, M., {et~al.} 2013, \aap, 556, A54

\bibitem[{{Whaley} {et~al.}(2009){Whaley}, {Irwin}, {Madden}, {Galliano}, \&
  {Bendo}}]{2009MNRAS.395...97W}
{Whaley}, C.~H., {Irwin}, J.~A., {Madden}, S.~C., {Galliano}, F., \& {Bendo},
  G.~J. 2009, \mnras, 395, 97

\bibitem[{{Williams} {et~al.}(2006){Williams}, {Borkowski}, {Reynolds},
  {Blair}, {Ghavamian}, {Hendrick}, {Long}, {Points}, {Raymond}, {Sankrit},
  {Smith}, \& {Winkler}}]{2006ApJ...652L..33W}
{Williams}, B.~J., {Borkowski}, K.~J., {Reynolds}, S.~P., {et~al.} 2006, \apjl,
  652, L33

\bibitem[{{Xilouris} {et~al.}(1998){Xilouris}, {Alton}, {Davies}, {Kylafis},
  {Papamastorakis}, \& {Trewhella}}]{1998A&A...331..894X}
{Xilouris}, E.~M., {Alton}, P.~B., {Davies}, J.~I., {et~al.} 1998, \aap, 331,
  894

\bibitem[{{Xilouris} {et~al.}(1999){Xilouris}, {Byun}, {Kylafis}, {Paleologou},
  \& {Papamastorakis}}]{1999A&A...344..868X}
{Xilouris}, E.~M., {Byun}, Y.~I., {Kylafis}, N.~D., {Paleologou}, E.~V., \&
  {Papamastorakis}, J. 1999, \aap, 344, 868

\bibitem[{{Xilouris} {et~al.}(1997){Xilouris}, {Kylafis}, {Papamastorakis},
  {Paleologou}, \& {Haerendel}}]{1997A&A...325..135X}
{Xilouris}, E.~M., {Kylafis}, N.~D., {Papamastorakis}, J., {Paleologou}, E.~V.,
  \& {Haerendel}, G. 1997, \aap, 325, 135

\bibitem[{{Yim} {et~al.}(2011){Yim}, {Wong}, {Howk}, \& {van der
  Hulst}}]{2011AJ....141...48Y}
{Yim}, K., {Wong}, T., {Howk}, J.~C., \& {van der Hulst}, J.~M. 2011, \aj, 141,
  48

\bibitem[{{Zaritsky}(1994)}]{1994AJ....108.1619Z}
{Zaritsky}, D. 1994, \aj, 108, 1619

\end{thebibliography}

\appendix

\section{A simple geometrical model}
\label{app:model}

For our analysis we considered strips of $-2.5', +2.5'$ from the centre of the galaxy and parallel to its major axis.
In this region the radial profile is relatively constant (emission always larger than $\sim 1/10$ of the peak) and 
we therefore considered the median emission to compute the vertical profile of the galaxy.
In this appendix we use a simple geometrical model to show the general validity of our approach, 
and study the bias in the estimates induced by geometrical components we did not account for in the fitting.
%we show that a more sophisticated geometrical model will lead to 
%very similar results, demonstrating the validity of our assumption.

As an example we consider PACS 100 $\mu$m data throughout this section.
We construct an array with a pixel size equal to the corresponding
instrument pixel size (i.e. $1''$) with flat null background. 
We perform a fit to the radial emission profile and 
%We perform a 6th degree polynomial fitting to the radial emission profile
%and 
assumed it as the radial profile of the intrinsic emission distribution.
We use the intrinsic vertical profile extracted in region X (see Sect.~\ref{sect:fitting},
$z_{\rm d,1} = 0.114$\,kpc and $z_{\rm d,2} = 1.02$\,kpc) as an estimate of the real
vertical profile and we added it to all the pixels around the midplane.
The simple model constructed in this way is then smoothed to the instrument
resolution convolving it to the corresponding PSF (the B(mod) PSF for PACS 100 $\mu$m) 
and adjusted in order to match the observed surface brightness.
\begin{table}[here]
\caption{Scaleheights of the thin and thick components as extracted (at galactic scale and in region X) from observations (obs.) and from our geometrical model.
Case $\alpha$ and case $\beta$ refer to the model without and with the inclusion of a super-thin disc, respectively.}             
\label{table:model_app}      
\centering          
\begin{tabular}{@{\extracolsep{1mm}} l c c c c}
\hline\hline
& \multicolumn{2}{c}{galactic} & \multicolumn{2}{c}{region X}  \\
\cline{2-3}
\cline{4-5}
& $z_{\rm d,1}$ &$z_{\rm d,2}$ & $z_{\rm d,1}$ & $z_{\rm d,2}$  \\
& (kpc) & (kpc) & (kpc) & (kpc) \\
\hline
obs. & $0.087 \pm 0.003$& $0.63 \pm 0.02$ & $0.114 \pm 0.004$& $1.02 \pm 0.10$\\
$\alpha$ & $0.113 \pm 0.004$& $0.98 \pm 0.06$ & $0.111 \pm 0.004$& $1.01 \pm 0.10$\\
$\beta$ & $0.069 \pm 0.002$& $0.87 \pm 0.06$ & $0.111 \pm 0.004$& $1.04 \pm 0.07$ \\
\hline
\end{tabular}
\end{table}

\begin{figure}[here]
\begin{center}
\includegraphics[trim=0mm 0mm 0mm 0mm,scale=0.9]{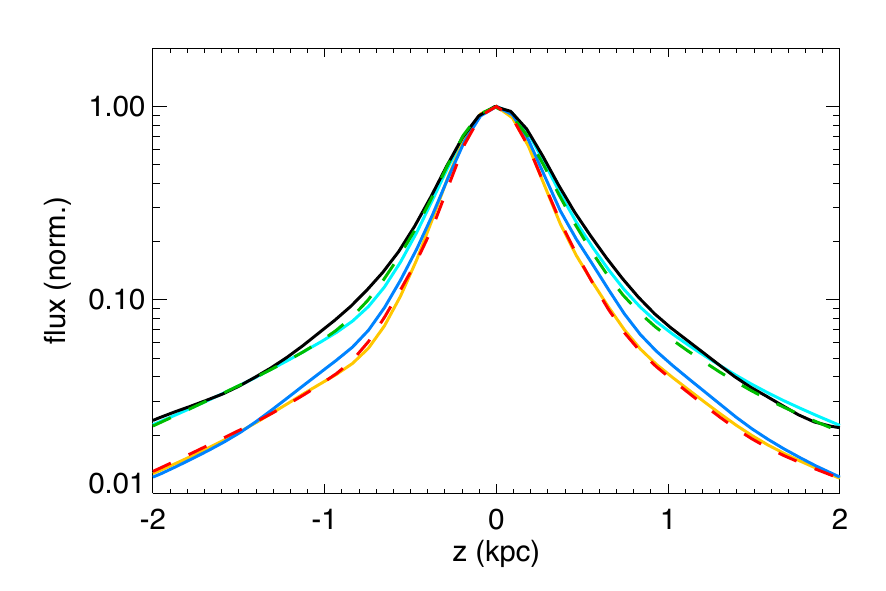}
\caption{Normalised vertical profiles for PACS 100 $\mu$m. Blue and black solid lines refer to the observed profiles at galactic scale and for region X, respectively.
Cyan and orange solid lines represent the vertical profiles obtained from our simple geometrical model without (case $\alpha$) and with (case $\beta$) a super-thin disc included, respectively,
while green and red dashed lines indicate the results of the fit in the two cases.}
\label{fig:model_app}
\end{center}
\end{figure}

We extract vertical profiles by averaging over the whole galactic disc and in region X.
In Fig.~\ref{fig:model_app} we show the vertical profile extracted in region X from our model
(cyan solid line) and compare it to the observed
vertical profile from region X (black solid line).
We then consider a two-component intrinsic exponential profile (Eq.~\ref{eq:vprof}),
convolve it with the instrument LSF and fit it to the vertical profile of the convolved model (likewise in Sect.~\ref{sect:fitting}).
The resulting profile is illustrated by the green dashed line in Fig.~\ref{fig:model_app} and
the parameters retrieved from the fit are reported as case $\alpha$ in Table~\ref{table:model_app}.
The scaleheight of the thin and thick components for both regions are compatible with those 
used to construct our simple geometrical model.
The compatibility between these results demonstrates that the assumption of a flat radial profile 
does not introduce any systematic error on the estimation of the vertical scaleheights.

%In Fig.~\ref{fig:model_app} we show the vertical profiles as observed at galactic scale (blue solid line)
%and in region X (black solid line).
%Also, the vertical profile extracted from our simple geometrical model (either at galactic scale or from region X) is shown
%in green dashed line.
%We note that the vertical profile from our simple model is very close to the observed vertical profile
%extracted in region X.
%The vertical scaleheights at each wavelength are compatible to 
%those retrieved in region X assuming a flat midplane emission and using LSFs.
%This demonstrates that our assumption is valid and that LSFs 
%together with radially constant profiles can be adopted without 
%affecting our results.

However, this simple model is not able to reproduce the discrepancy in
scaleheight fitting observations at galactic scale or considering only region X,
which could be due to unresolved sources not accounted for in our fitting procedure (see Sect.~\ref{sect:vert_var} and \ref{sect:thincomp}).
To test this hypothesis, we made two simple tests. 

In the first test, we include a few (5) isolated point sources along the galactic plane, 
to simulate the main features in the observed surface brightness distribution. 
These features could represent inhomogeneities in the disc due to e.g. spiral arms seen 
in projection (see the discussion in \citealt{2011A&A...531L..11B}). 
Being point-like, their vertical profile would follow that of a PSF, and this would make the 
average profile decline faster than in the case of two diffuse discs. 
The intensity of the point sources (modelled as bright pixels) and of the diffuse discs 
where scaled to match the observed average profile, and an average profile was extracted 
from the model as we have done before. Though the new profile (not shown) is narrower 
than the profile for region X, it is very close to it and far from the observed averaged profile. 
Thus, isolated point sources are not likely to be the main cause of the discrepancy.

In a second test, we assume that half of the emission in the midplane comes 
from a very thin diffuse disc which is not resolved vertically 
(and thus has a vertical profile similar to that of the LSF).
%In order to test this hypothesis, we assume that half of the emission in the midplane
%actually comes from unresolved sources which are not vertically extended.
This is analogous to dividing the intrinsic profile by a factor two for all the pixels 
which are not in the midplane.
An exception is made for the pixels around region X where no bright source is added.
%In order to test this hypothesis, at the position of the main peaks in the midplane
%(blue dotted lines in Fig.~\ref{fig:radial_multi}), we add one bright pixel to the intrinsic
%distribution. 
We then convolve this model with the instrument PSF and adjust the normalisation
to the observed surface brightness.
The vertical profile at galactic scale of the convolved model is shown 
by the orange solid line in Fig.~\ref{fig:model_app} and compared to the observed vertical profile at galactic scale (blue solid line).
With the same method used above, the scaleheight of the thin and thick components 
are retrieved from fitting our convolved model. 
The resulting vertical profile is shown in Fig.~\ref{fig:model_app} in red dashed line 
and fitting parameters are reported in Table~\ref{table:model_app} (case $\beta$).
%The vertical profile extracted at galactic scale from our convolved model is shown in Fig.~\ref{fig:model_app}
%(red dashed line) 
The inclusion of an unresolved disc reduces the scaleheight of both the thin and thick components 
leading to a good agreement between the convolved and observed vertical profiles at
galactic scale.
Thus, the presence of a super-thin disc can explain the discrepancy between the vertical 
profiles extracted from observations at galactic scale and in region X.

It has to be noted that the scaleheight of the thin and thick components retrieved from fitting
our convolved model in region X are not affected by the inclusion of unresolved sources
along the midplane (see Table~\ref{table:model_app}; case $\beta$, region X).
However, this does not imply that we can exclude that part of the emission from the midplane of region X 
is actually due to the presence of unresolved sources. In fact, this represents a scenario that we are not able to 
test with this simple geometrical model.

%\begin{tabular}{@{\extracolsep{1mm}} l c c c c c c }
%\hline\hline
%& \multicolumn{2}{c}{IRAC 8 \mic} & \multicolumn{2}{c}{MIPS 24 \mic} & \multicolumn{2}{c}{PACS 70 \mic}  \\
%\cline{2-3}
%\cline{4-5}

%The vertical scaleheights of both the dust components are systematically $\sim 6-7\%$ 
%lower than, and not compatible to, those obtained without including bright point sources to the intrinsic distribution.
%This indicates that the presence of point sources affects both scaleheights 
%at galactic scale, reducing them to values closer to those observed.

%Io qui metterei Appendix B: Beyond "standard" Herschel PSFs. Poi un cappellino 
%per tutti. In this work we attempted to take into account the major effects effects that 
%could modify the PSF of PACS and SPIRE (dire perche' non consideriamo MIPS/IRAC). 
%In particular, we studied the variations induced to colour corrections (B.1), we 
%derived a new set of "observed" PSFs for PACS (B.2) and finally tested the impact of these 
%modifications on the derivation of the vertical scalelelngths (B.3) 
%
%e poi le chiami come sottosezioni, cosi' abbiamo solo "due" appendici!

\section{Beyond ``standard'' \textit{\textbf{Herschel}} PSFs.}

In this work we attempt to take into account the major effects that 
could modify the PACS and SPIRE PSFs.
In particular, we study the variations induced by colour corrections (B.1), we 
derived a new set of ``observed'' PSFs for PACS (B.2) and finally tested the impact of these 
modifications on the derived vertical scaleheights (B.3). 

\subsection{PSF modification according to the source spectrum}
\label{app:PSF_mod}

The effective frequency, $\nu_{\rm eff}$, of an instrument equipped with a narrowband filter 
corresponds to the frequency at the centre of the filter.
For larger filters, $\nu_{\rm eff}$ depends on the spectrum of the observed source and 
therefore the instrument PSF shrinks or stretch according to the source spectrum.
In the latter case, a PSF estimated from observations of an astronomical object is only 
valid if the source that we observe has a similar spectrum to that of the 
object used for the PSF characterisation.
However, the spectrum of an asteroid or a planet can be very different from the dust 
emission spectrum.
For this reason we modified the PACS and SPIRE PSFs according to the typical dust temperature
and $\beta$ values.

Following the SPIRE Handbook (2014) 
a PSF depends on the frequency of the observed source as:
\begin{equation}
B(\theta,\nu) = B(\theta \times \left( \frac{\nu}{\nu_{\rm eff}}\right)^{\gamma},\nu_{\rm eff}),
\end{equation}
where $\theta$ is the radial distance from the centre, $\nu$ is the source frequency, 
$\nu_{\rm eff}$ is the effective frequency for which the PSF has been estimated
and $\gamma$ is a parameter which takes into account the slope of the spectrum
within a given band ($\gamma = -1$ for an ideal telescope).

The SPIRE ICC provides values of $\gamma$ and $\nu_{\rm eff}$ for all SPIRE bands, 
in particular $\gamma = -0.85$ and $\nu_{\rm eff} = 1215$\,GHz for SPIRE 250 $\mu$m.

\begin{figure}[here]
\begin{center}
\includegraphics[trim=0mm 0mm 0mm 0mm,scale=0.9]{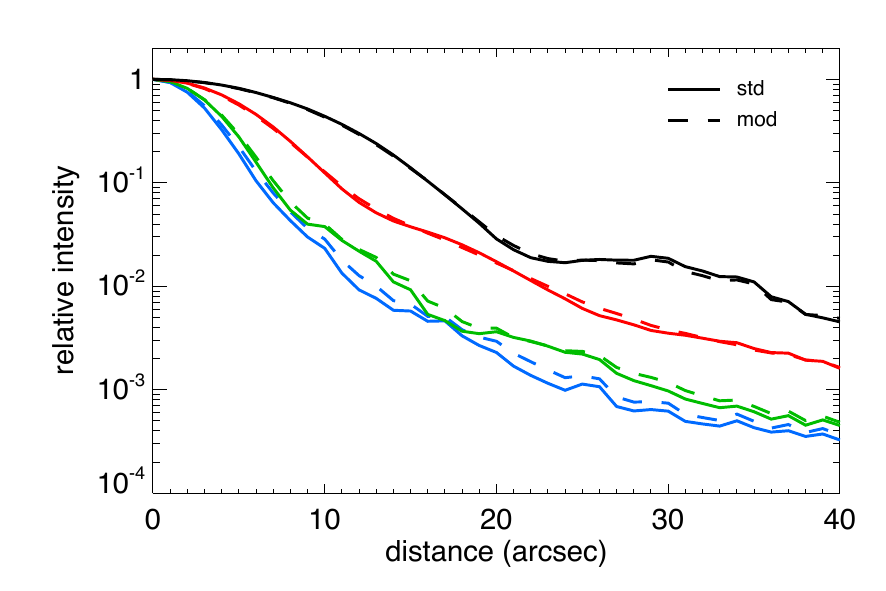}
\caption{PACS  70, 100 and 160 $\mu$m (blue, green and red lines, respectively) and 
SPIRE 250 $\mu$m (black lines) PSFs, standard and modified assuming 
$T_{\rm dust} = 20$\,K and $\beta_{\rm dust} = 2$.}
\label{fig:PSFs_mod_std}
\end{center}
\end{figure}
For PACS bands we assume $\gamma = -1$ and we estimate $\nu_{\rm eff}$
from the spectrum of Vesta and Mars as follows:
\begin{equation}
\nu_{\rm eff} = \frac{\int_0^{\infty} \nu S(\nu) T(\nu) W(\nu) d\nu}{\int_0^{\infty}S(\nu) T(\nu) W(\nu) d\nu},
\end{equation}
where $S(\nu)$ is the source of the object used for the estimation of the PSF, $T(\nu)$ is the
transmission function and $W(\nu)$ the bolometer response\footnote{PACS and SPIRE filter transmission functions 
and bolometer response can be retrieved can be accessed from the HIPE environment.}.

Finally we compute the PSF for a given spectrum, $S'(\nu)$ of the object that we want to observe (i.e. dust typically
at $T = 20$\,K and $\beta \sim 2$):
\begin{equation}
B(\theta,T, \beta) = \frac{\int_0^{\infty} T(\nu) S'(\nu) W(\nu) B(\theta,\nu) d\nu}{\int_0^{\infty} T(\nu) S'(\nu) W(\nu) d\nu}.
\end{equation}
For a more complete description of the method used the reader is referred to \cite{mythesis}.

In Fig.~\ref{fig:PSFs_mod_std} we show PACS 70, 100 and 160 $\mu$m and SPIRE 250 $\mu$m PSFs
for the standard case and modified according to the typical dust emission spectrum.
Here, the ``standard'' PSFs for PACS are B (std) while for SPIRE are A (std).
We note that for all wavelengths variations are never more relevant than $1\%$ level.

Given the lack of information on the IRAC and MIPS monochromatic PSFs and given the low 
impact of the colour corrections for PACS and SPIRE PSFs we do not perform this work for
{\it Spitzer} PSFs.

\subsection{PACS PSFs estimation from observations}
\label{app:psfobs}

\begin{figure}[here]
\begin{center}
\includegraphics[trim=0mm 0mm 0mm 0mm,scale=0.9]{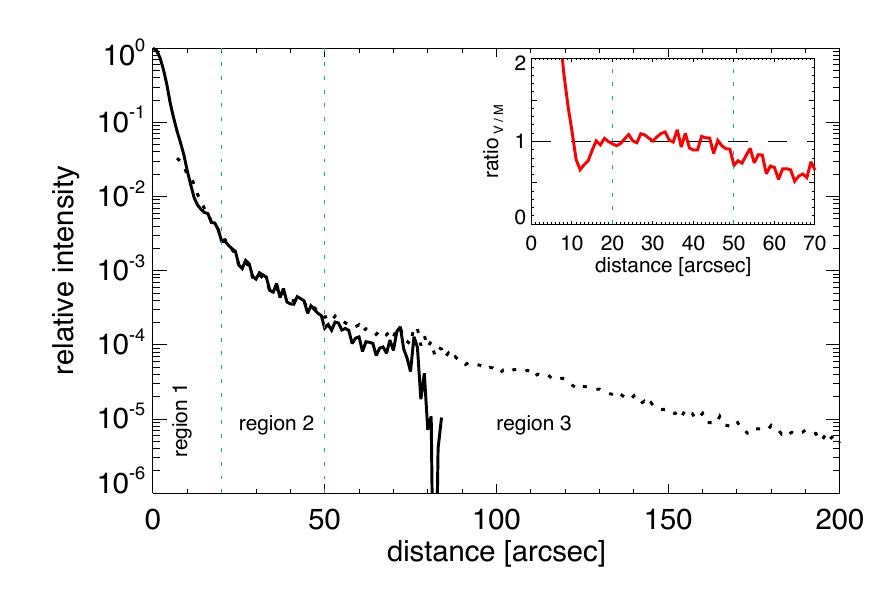}
\caption{Radial profiles of Vesta (black solid line) and Mars (black dotted line) observations at 70 $\mu$m. In the inset the ratio between 
the Vesta and Mars profiles is shown (red solid line). Regions are indicated (see text for details).}
\label{fig:profile_combine}
\end{center}
\end{figure}

In order to obtain a good estimate of an instrument PSF it is necessary to have
information both on its core and on faint structures in the outskirts.
However, this involves a wide range in intensity and observations of a single object
are often not sufficient to properly characterise a PSF. 

The PACS ICC provides an optical model for the instrument PSF taking into account aberrations of
the real telescope (``as built'' PSF\footnote{\url{http://herschel.esac.esa.int/twiki/pub/Public/PacsCalibrationWeb/PACSPSF_PICC-ME-TN-029_v2.0.pdf}}).

We use observations of an asteroid, Vesta, and a planet, Mars, to compare 
the observed PSF with that modeled by the ICC.
Vesta is a sufficiently faint object not to lead to any saturation or non-linear 
behaviour of the central pixels, therefore allowing a good estimate of the
core of the PSF.
On the other hand, Mars is much brighter and leads to the complete saturation
of the central pixels but gives much more information on the PSF faint wings.
The combination of these observations can thus lead to an accurate estimate of the PSF.
\begin{figure}[here]
\begin{center}
\includegraphics[trim=0mm 0mm 0mm 0mm,scale=0.9]{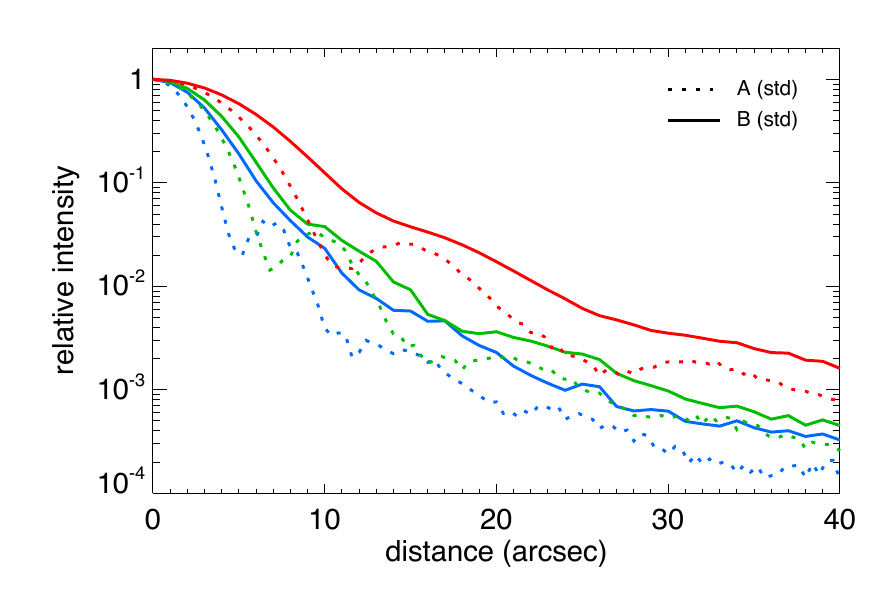}
\caption{PSF Radial profiles as estimated from observations (solid lines) and modeled by the ICC (dotted lines) for
PACS 70 $\mu$m (blue lines), 100 $\mu$m (green lines) and 160 $\mu$m (red lines).}
\label{fig:PACS_ICC_comb}
\end{center}
\end{figure}

Observations of Vesta (ObsID: 1342195472, 1342195473 for 70 \mic, 
1342195476, 1342195477 for 100 \mic\,and 1342195472, 1342195473, 1342195476, 1342195477 for 160 \mic) and Mars 
(ObsID: 1342231157 - 1342231160 for 70 \mic, 1342231161 - 1342231164 for 100 \mic\,and 1342231157 - 1342231164 for 160 \mic) 
were reduced following the same method as described in Section~\ref{sect:obs}.

First, all the images have been rotated in a way to obtain the Z-axis of the telescope during observations
pointing upward. 
Then, for all the images we removed an average background flux computed 
in regions sufficiently far from the source.
We normalised the images of Vesta to unity and rescaled the images of Mars so to
have the same flux in an intermediate region between the core and the outskirts of the PSF.

The average radial profiles of Vesta and Mars for PACS 70 $\mu$m are shown in Fig.~\ref{fig:profile_combine}.
The inset illustrates the ratio between the flux of Vesta and Mars as a function of the distance from the PSF centre.
We note in a region close to the centre (region 1), the radial profile extracted from Mars does not follow that of Vesta
because of saturation / non-linear issues, while at distances $\gtrsim 70''$ (region 3) the Vesta profile 
is not detected at a significant level.
At intermediate distances (region2, from $x_1 \sim 20''$ to $x_2 \sim 50''$) the ratio is close to unity.
As an estimate of the PSF we use Vesta observations in region 1, Mars observations in region 3 and the 
average value between the two for region 2.
We performed the same operation for PACS 100 and 160 $\mu$m, using $x_1 = 30$ and $x_2 = 50$ for PACS 100 $\mu$m
and $x_1 = 40$ and $x_2 = 70$ for PACS 160 $\mu$m.

In order to obtain a smooth PSF in $x_1$ and $x_2$ we adopt the following smooth step function (\citealt{smooth}):
\begin{equation}
f(x) = 6 \left( \frac{x-x_1}{x_2-x_1}\right)^5 - 15 \left(\frac{x-x_1}{x_2-x_1}\right)^4 +10 \left(\frac{x-x_1}{x_2-x_1}\right)^3,
\end{equation}
which has zero first and second order derivatives at the boundaries $x_1$ and $x_2$.

In Fig.~\ref{fig:PACS_ICC_comb} we show radial profiles of the PACS PSFs as released by
the ICC and obtained in this way.
Differences are evident at all wavelengths and distances.
In particular, ICC PSFs are systematically narrower by 10-30\% than those we obtained from observations.

\subsection{Scaleheights assuming different PSFs}
\label{app:plots_scaleheights}

\begin{figure}[here]
\begin{center}
\includegraphics[trim=0mm 0mm 0mm 0mm,scale=0.9]{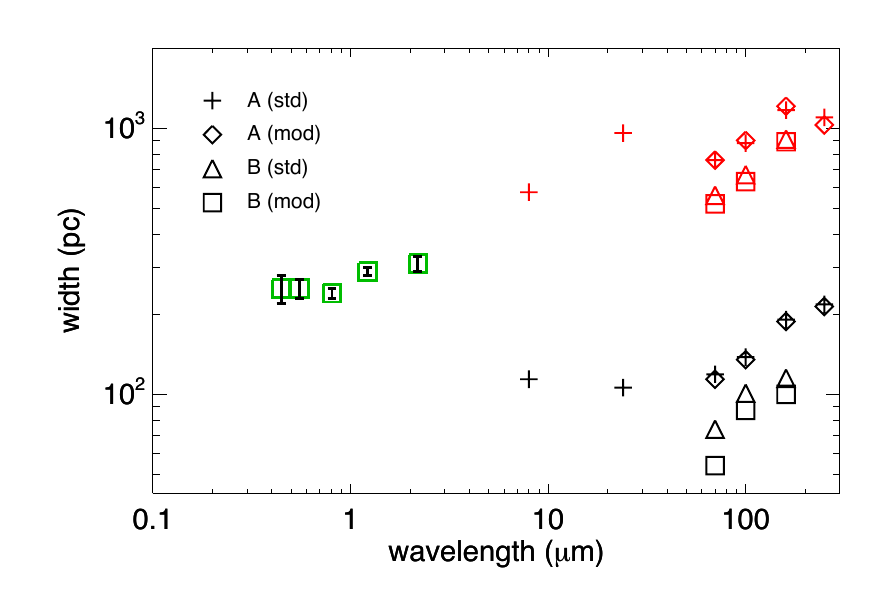}
\caption{Scaleheights of the thin (black symbols) and thick (red symbols) dust components assuming different PSFs. 
Green squares represent the dust scaleheights as estimated by \cite{1999A&A...344..868X}.}
\label{fig:widths_appendix}
\end{center}
\end{figure}

With the method described in Section~\ref{sect:dust_em} we estimated
the scaleheight of the dust emission profile at different wavelengths.
The instrument PSF plays a key role in this calculation.
In Fig.~\ref{fig:widths_appendix} we show the scaleheights as retrieved 
from the fitting procedure using the different PSFs we considered in the main text.
We note that the effect of modifying the PSF according to the observed source
spectrum has no relevant effect on the estimation of the scaleheight, reflecting
the small effect on the PSF (see Section~\ref{app:PSF_mod}).
On the contrary, passing from the ``as built'' PACS PSF (A) to that estimated from observations (B)
importantly affects the dust scaleheight.

% the polynomial fit is justified because as already observed for example
% by Bianchi & Xilouris there is a break in the radial profile.

%Then, at the position of the main peaks in the midplane
%(blue dotted lines in Fig.~\ref{fig:radial_multi}), we add one bright pixel. 
%The simple model constructed in this way is then smoothed to the instrument
%resolution convolving it to the corresponding PSF. We adjust the
%normalisation of the vertical profile and that of each point source
%in order to match the observed surface brightness for most of the pixels in
%the midplane.

%We extract vertical profiles at galactic scale and for each of
%the regions described earlier. The vertical profile variability as well as
%the vertical profile at galactic scale are very similar (i.e. the discrepancy between scaleheights
%is $< 20\%$)
%to those obtained under the assumption of constant radial emission between $-2.5', +2.5'$ from the
%galactic centre.
%This demonstrates that our assumption is valid and that LSFs 
%together with radially constant profiles can be adopted without 
%affecting our results.

%In particular, the width of the
%vertical profile in region X is equal to that retrieved from observations,
%indicating that this region is not affected by the point
%sources present in the midplane. This makes the vertical profile
%extracted from this region the best estimate of the real vertical
%profile. Scaleheights of the thin (black dots) and thick (red dots)
%discs retrieved from fitting observations in region X are indicated
%in Fig. 3 and listed in Table 1.

\end{document}